\documentclass[12pt]{article}

\usepackage[letterpaper,margin=3cm]{geometry}

\usepackage{authblk}

\usepackage[parfill]{parskip}
\usepackage{setspace}

\usepackage[labelfont=bf,font={small,stretch=0.95}]{caption}

\usepackage[super,sort,compress,numbers]{natbib}

\usepackage{gensymb}
\usepackage{amsmath}
\usepackage{amssymb}
\usepackage{mathtools}
\usepackage{comment}
\usepackage{hyperref}
\usepackage{threeparttable}
\usepackage{placeins}

\usepackage{graphicx}
\usepackage{multirow}

\usepackage[version=4]{mhchem}

\usepackage{mathptmx}

\usepackage{xcolor}
\usepackage{soul}
\usepackage{booktabs}

\begin{document}
\title{\fontsize{20}{26}\selectfont\bf Amorphous materials as a frontier challenge for universal interatomic potentials}

\author[1]{Natascia L. Fragapane}
\author[1]{Volker L. Deringer\thanks{volker.deringer@chem.ox.ac.uk}}

\affil[1]{Inorganic Chemistry Laboratory, Department of Chemistry, University of Oxford,\protect\\ Oxford, UK}

\date{}

\maketitle

\setstretch{1.5}

{\bf
Pre-trained or `foundational' machine-learned interatomic potentials (MLIPs) are now widely used in materials modelling. However, early pre-trained models and benchmarks have largely focused on ordered, crystalline structures, and their transferability to non-crystalline solids remains unclear. Here, we show that the amorphous state is indeed a central challenge for future universal MLIPs, based on a systematic evaluation of current mainstream models in this domain. We introduce a benchmarking framework built on a curated reference dataset of canonical amorphous systems, as well as validation for structures and properties. Our study identifies limitations in the transferability of many current pre-trained models and investigates fine-tuning strategies tailored to disordered phases. Together, our results can facilitate future applications of MLIPs in the fast-growing field of amorphous functional materials, and they provide guidance for designing next-generation training datasets and transferable atomistic models.
}

\clearpage

Machine-learned interatomic potentials (MLIPs) have evolved from specialist tools to widely accessible components of the atomistic modelling toolkit \cite{deringer_machine_2019, unke_machine_2021, friederich_machine-learned_2021}. Central to this shift is the emergence of pre-trained interatomic potentials trained on extensive datasets that span diverse chemical elements and materials systems \cite{yang_mattersim_2024, park_scalable_2024, neumann_orb_2024, rhodes_orb-v3_2025, batatia_foundation_2025, batatia_cross_2025, kim_optimizing_2025, mazitov_pet-mad_2025, tan_high-performance_2025, zhou_matris_2026, lysogorskiy_graph_2026, wood_uma_2026}. These models provide practical starting points for simulations across broad chemical space and can often be refined for specific applications through targeted fine-tuning. These advances have in turn motivated the long-term goal of developing truly `universal' MLIPs, capable of reliable out-of-the-box performance across diverse chemistries and simulation regimes.

The strong performance of current pre-trained MLIPs is closely tied to the predominance of ordered materials in both training datasets and evaluation benchmarks \cite{deng_chgnet_2023, schmidt_improving_2024, barros-luque_open_2026, kaplan_foundational_2025, levine_open_2026, gharakhanyan_open_2026}. Early pre-training datasets were constructed primarily from relaxation trajectories of structures from the Materials Project \cite{jain_commentary_2013, horton_accelerated_2025}, providing good coverage of crystalline configurations near local and global energetic minima. More recent datasets, including Alexandria \cite{schmidt_improving_2024}, OMat24 \cite{barros-luque_open_2026}, MATPES \cite{kaplan_foundational_2025}, and MAD \cite{mazitov_massive_2025, malosso_high-quality_2026}, have expanded sampling by incorporating a wider range of hypothetical crystals, high-temperature molecular-dynamics (MD) trajectories, and systematically perturbed structures. Together, these developments have greatly improved the coverage of crystalline regions of configurational space, enabling models trained on such datasets to achieve increasingly accurate zero-shot results for ordered structures.

However, the requirements for universal MLIPs reach beyond crystalline domains. Many technologically important materials -- including recent examples in energy storage \cite{zhang_family_2023}, phase-change memory \cite{wang_amorphous_2026}, and photocatalysis \cite{wang_induced_2024} -- have disordered or fully amorphous structures that are directly linked to properties and applications \cite{liu_amorphous_2024}. The absence of long-range structural order and the increased diversity of local bonding motifs make the amorphous state a critical test case for the generality of pre-trained MLIPs.

Despite increasing interest in amorphous materials, the systematic evaluation of pre-trained MLIPs in this domain remains limited. Current general-purpose materials benchmarks, including MatBench \cite{dunn_benchmarking_2020, riebesell_framework_2025}, JARVIS \cite{choudhary_jarvis-leaderboard_2024}, LAMBench \cite{peng_lambench_2026}, and the recently introduced Dyna-Mat \cite{gawkowski_dyna-mat_2026}, are primarily orientated towards ordered systems. More targeted studies have begun to address this gap: the original MACE-MP-0 work \cite{batatia_foundation_2025} reported tests on amorphous carbon and silicon, while EGraffBench \cite{bihani_egraffbench_2024} and later LiPS-25 \cite{fragapane_lips_2026} include disordered Li--P--S configurations in their numerical benchmarks. More recently, pre-trained models have also been assessed for amorphous structure generation with melt--quench simulations, both for glassy solid-state electrolytes \cite{bertani_atomic_2025, bertani_machine_2026} and aluminosilicate glasses \cite{benassi_assessing_2026}, and across a broad range of chemistries \cite{li_melt-quench_2026}, where systematic failures of leading pre-trained models were identified. However, these studies address specific aspects of amorphous materials modelling, rather than providing a unified benchmarking framework. As a result, it is unclear how far today's mainstream MLIPs can be considered truly `universal'.

Here, we systematically assess a range of pre-trained MLIPs in the domain of amorphous materials. We first evaluate their zero-shot numerical accuracy on AM26, a curated benchmark dataset comprising five prototypical and well-understood amorphous systems: from elements to ternary compounds. We then assess the quality of amorphous structures generated using these MLIPs, as well as the models' ability to reproduce characteristic structural transformations. Finally, we investigate the extent to which fine-tuning can improve performance in this regime. By revealing where existing models succeed and where they fall short, this work provides guidance for the application of pre-trained MLIPs to amorphous materials and informs the design of next-generation training datasets that better capture structural disorder.

\section*{Results}
\subsection*{The AM26 dataset and benchmarks}
\begin{figure*}[t]
    \centering
    \includegraphics[width=\linewidth]{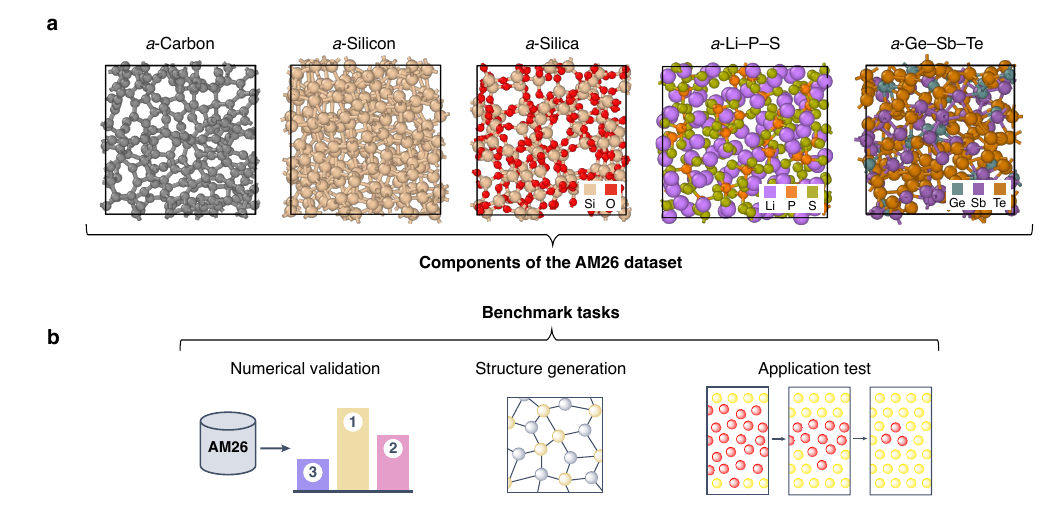}
    \caption{\textbf{Benchmarking MLIPs for amorphous materials.} 
    (\textbf{a}) Examples of structures from the AM26 dataset, spanning five prototypical bulk amorphous systems. Structures were visualised with OVITO \cite{stukowski_visualization_2009}.
    (\textbf{b}) Schematic overview of the benchmarking protocol, comprising three tasks: (i) numerical validation, quantifying energy and force accuracy on the AM26 dataset; (ii) structure generation, assessing the quality of MLIP-generated amorphous structures under standardised simulation protocols for three key systems (\textit{a}-C, \textit{a}-\ce{SiO2}, and \textit{a}-GST); and (iii) application tests, probing the reliability of larger-scale simulations characteristic of amorphous materials.} 
    \label{fig:figure_1}
\end{figure*}

Our benchmarking framework for assessing MLIPs in the context of amorphous materials involves three categories of tasks (Fig.~\ref{fig:figure_1}): (i) numerical accuracy; (ii) the generation of realistic structures; and (iii) the performance in downstream simulations characteristic of real-world amorphous materials modelling. Underpinning these tasks, we introduce a curated dataset of first-principles data, which we call AM26. The AM26 dataset spans five representative and well-studied bulk amorphous (\textit{a}-) systems -- \textit{a}-C, \textit{a}-Si, \textit{a}-\ce{SiO2}, \textit{a}-Li--P--S (`$a$-LiPS'), and \textit{a}-Ge--Sb--Te (`$a$-GST') -- ranging from from complex elemental structures to multi-component ternary glasses, and anchoring the benchmark in materials systems for which previously-validated, specialist MLIPs exist. Detailed benchmarking protocols and analysis procedures are provided in the Supplementary Information.

The configurations in AM26 were generated using established melt--quench protocols with existing MLIPs for \textit{a}-C, \textit{a}-\ce{SiO2}, and \textit{a}-LiPS \cite{deringer_machine_2017, erhard_machine-learned_2022, erhard_modelling_2024, fragapane_lips_2026}; we further added previously reported structures for \textit{a}-Si \cite{rosset_signatures_2025} and \textit{a}-GST \cite{zhou_device-scale_2023}. Unlike datasets designed primarily for MLIP training, which may include very-high-energy structures to improve model robustness, AM26 focuses on physically meaningful configurations representative of the amorphous state. All structures in AM26 were (re-) labelled with density-functional theory (DFT) energies and forces computed using the \texttt{MPStaticSet} workflow \cite{jain_commentary_2013, ong_python_2013}, consistent with recent pre-training datasets such as OMat24 \cite{barros-luque_open_2026} (see Methods section for details). The full dataset is openly available at Ref.~\citenum{fragapane_am26_2026}. Future extensions to additional amorphous material systems are anticipated.

\subsection*{Benchmarking pre-trained MLIPs}
\begin{figure*}[htbp]
    \centering
    \includegraphics[width=\linewidth]{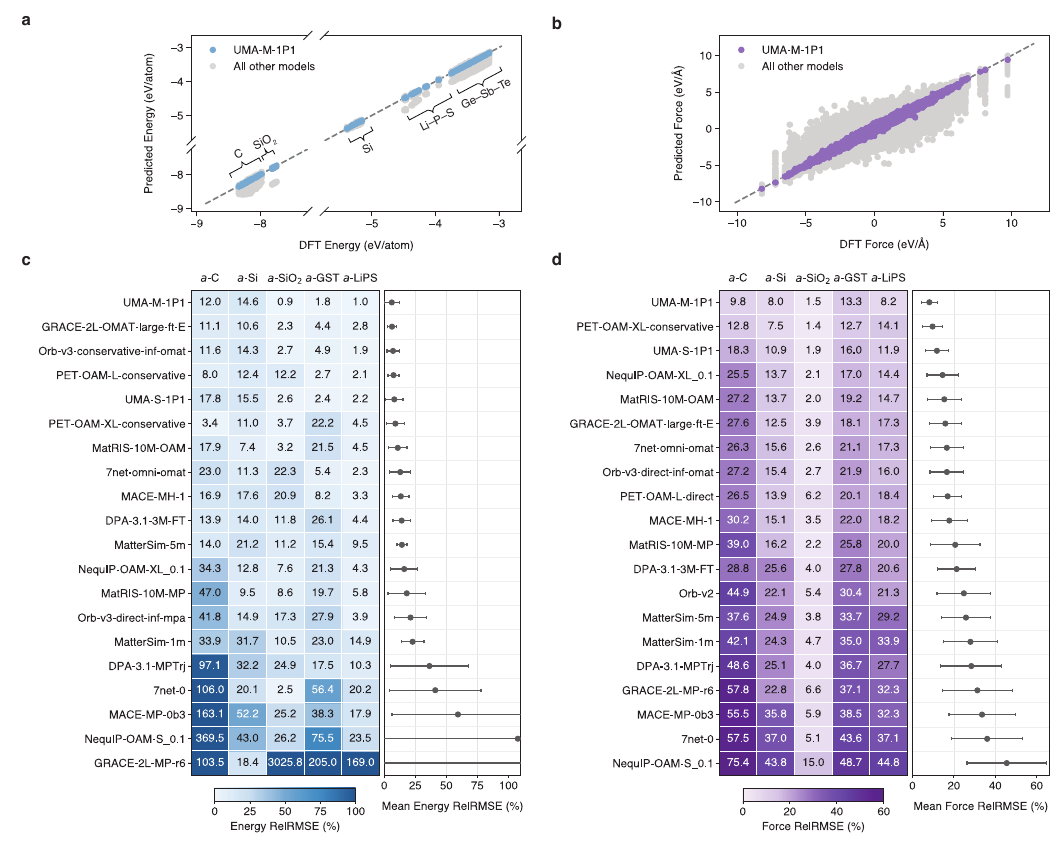}
    \caption{\textbf{Benchmarking pre-trained MLIPs on the AM26 dataset.} We show parity plots comparing (\textbf{a}) energy and (\textbf{b}) force predictions against DFT reference data across all AM26 subsystems. Predictions from the best-performing model (\texttt{UMA-M-1P1}) are highlighted in blue/purple; results for all others are shown in grey. In panel (a), labels identify the contributions from each AM26 subsystem. Relative root-mean-square errors (RelRMSEs), defined as the RMSE normalised by the standard deviation of the reference labels, are reported for (\textbf{c}) energies and (\textbf{d}) forces. Heatmaps (\textit{left}) display errors for each of the five AM26 subsystems, while dot plots (\textit{right}) report the corresponding mean RelRMSE across those five systems; models are ranked according to this value, and error bars indicate the standard deviation across systems. Only the best- and worst-performing models of each family are displayed here; more comprehensive results are provided in the Supplementary Information.} 
    \label{fig:figure_2}
\end{figure*}

Figure \ref{fig:figure_2} presents the AM26 leaderboard, summarising the numerical accuracy of pre-trained MLIPs in the domain of amorphous materials. Panels (a) and (b) compare MLIP-predicted energies and forces to the DFT reference labels of AM26. Panels (c) and (d) summarise the subsystem-resolved performance using heatmaps and mean performance across the five AM26 subsystems using dot plots. Errors are reported as relative root-mean-square errors (RelRMSEs), calculated by normalising the RMSE by the standard deviation of the corresponding labels: this metric enables comparison across systems with differing variability in their energy and force distributions, and thus differing levels of task difficulty. 

Figure \ref{fig:figure_2} characterises 20 representative pre-trained MLIPs: we selected the best- and worst- performing ones from each of 10 model families (GRACE \cite{lysogorskiy_graph_2026}, MACE \cite{batatia_foundation_2025, batatia_cross_2025}, MatterSim \cite{yang_mattersim_2024}, Mat\-RIS \cite{zhou_matris_2026}, NequIP \cite{tan_high-performance_2025}, Orb \cite{neumann_orb_2024, rhodes_orb-v3_2025}, PET-OAM \cite{mazitov_pet-mad_2025}, SevenNet \cite{park_scalable_2024}, DeePMD \cite{zhang_graph_2026}, and UMA \cite{wood_uma_2026}). An extended benchmark of 41 models is shown in the Supplementary Information. Most models were trained directly at the PBE(+U) level; others employ multi-fidelity architectures with separate output heads specialised to different DFT approximations. For multi-head models, the default output head was used unless otherwise specified, in which case the selected head is indicated explicitly in the model name (e.g., \texttt{7net-omni-omat}). Further details of pre-training data, model architectures, and training strategies are given in the Methods section. 

Across AM26, models that perform well overall tend to do so consistently between subsystems. Conversely, poorer-scoring models are not uniformly inaccurate: their overall error is typically dominated by failure on one or more particularly challenging cases. Indeed, the AM26 benchmark reveals substantial variation in difficulty across subsystems. The \textit{a}-\ce{SiO2} force predictions are relatively consistent across models, with only a $\sim$14\% spread between the best and worst performers included in Fig.~\ref{fig:figure_2}, suggesting that these structures -- consisting, by and large, of \ce{[SiO4]} tetrahedra similar to those in crystalline silicates -- are well described by current pre-trained MLIPs. In contrast, \textit{a}-C presents a much greater challenge: we find multiple RelRMSE values exceeding 100\% in energies and 50\% in forces. While the MPTrj dataset, underpinning many current pre-trained models, does contain carbon structures with sp$^2$ and sp$^3$ coexistence \cite{deng_chgnet_2023, batatia_foundation_2025}, these appear insufficient to capture the full density-dependent bonding and topological diversity present in realistic models of amorphous carbon \cite{caro_machine_2020}. 

Encouragingly, several models demonstrate strong and consistent performance across all AM26 subsystems. In particular, \texttt{UMA-M-1P1} \cite{wood_uma_2026} achieves the lowest average errors in both energies and forces, maintaining RelRMSEs below 15\% across all five datasets, consistent with the tight parity plots observed in Fig.~\ref{fig:figure_2}a--b. Strong zero-shot performance is not restricted to a single architecture, however: models from several distinct families achieve RelRMSE values below $\sim$20\% across all AM26 subsystems.
In contrast, some models exhibit severe failure modes -- particularly in energy prediction, where RelRMSE values occasionally exceed 100\%, indicating prediction errors larger than the intrinsic variability of the reference data. In the most extreme case we observed here, \texttt{GRACE-2L-MP-r6} \cite{lysogorskiy_graph_2026} yields an energy RelRMSE of $\sim$3000\% for \textit{a}-\ce{SiO2}. This does not seem to be a shortcoming of the architecture, given that \texttt{GRACE-2L-OMAT-large-ft-E} performs very well (energy RelRMSE of 2.3\% for \textit{a}-\ce{SiO2}, and second-best performance across all models in Fig.~\ref{fig:figure_2}c). Rather, this example suggests that pre-training strategy and data play a greater role in determining amorphous transferability than architecture alone.

A wider-ranging analysis across different model families is presented in Supplementary Fig.~1. Collectively, our results reveal substantial variability in the transferability of current pre-trained MLIPs -- suggesting that success on existing crystalline-focused evaluation frameworks does not guarantee accurate zero-shot performance in amorphous systems.

\begin{figure*}[htbp]
    \centering
    \includegraphics[width=\linewidth]{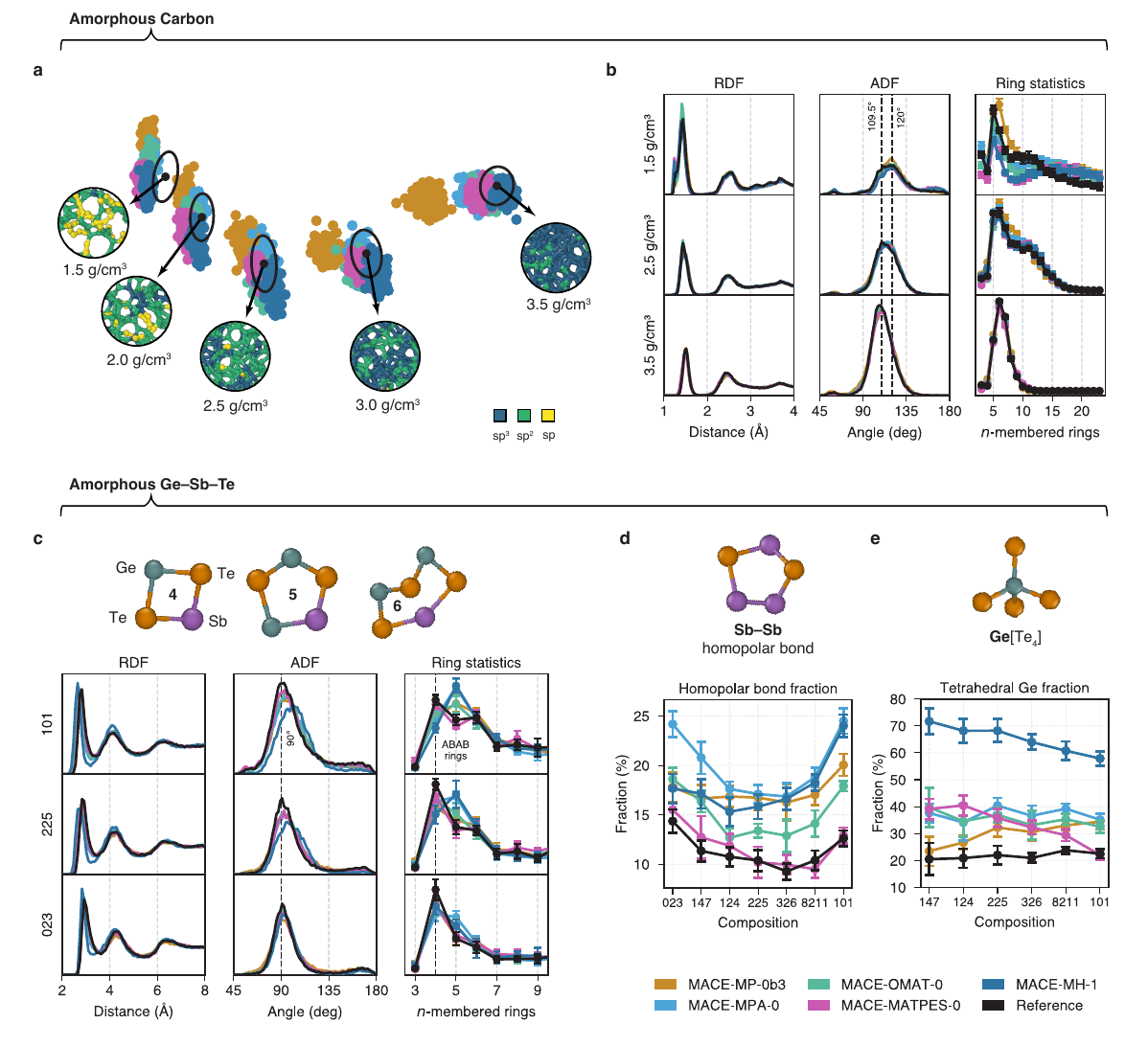}
    \caption{\textbf{Fidelity of amorphous structures generated by pre-trained MLIPs.}
    (\textbf{a}) A similarity map of $a$-C structures generated by different models, projected onto a principal component analysis (PCA) map of SOAP descriptors derived from \texttt{C-GAP-17} reference structures. Reference density clusters are represented by 95\% confidence ellipses; structures falling within these regions are considered structurally consistent with the reference ensemble. Representative structural snapshots are shown.
    (\textbf{b}) Radial distribution functions (RDFs), angle distribution functions (ADFs), and ring statistics for $a$-C structures across the density range 1.5--3.5 g/cm$^3$.
    (\textbf{c}) RDFs, ADFs, and ring statistics for $a$-GST structures along the GeTe--\ce{Sb2Te3} tie line. 
    (\textbf{d}) Fraction of homopolar bonds in amorphous GST. 
    (\textbf{e}) Fraction of tetrahedrally coordinated Ge atoms in amorphous GST.
    In panels (c--e), representative examples are shown above the plots (Ge: turquoise, Sb: dark purple, Te: dark orange). For panels (b) and (c), values represent averages over ten independently generated structures per density/composition; RDFs and ADFs were computed using 512-atom ($a$-C) and $\sim$500-atom ($a$-GST) structures; ring statistics were evaluated using 4096-atom ($a$-C) and $\sim$500-atom ($a$-GST) structures. For panels (d) and (e), values are averaged over the final 40~ps of ten independent melt-quench simulations. Error bars denote standard deviations.} 
    \label{fig:figure_3}
\end{figure*}

\subsection*{Structures}

We next assess whether pre-trained MLIPs can generate realistic amorphous structures in standard melt--quench simulations. 
We benchmark the quality of $a$-C, $a$-\ce{SiO2}, and $a$-GST structures generated by pre-trained MLIPs against those obtained with established specialist MLIPs: \texttt{C-GAP-17} \cite{deringer_machine_2017}, \texttt{SiO$_x$-ACE-24} \cite{erhard_modelling_2024}, and \texttt{GST-ACE-24} \cite{zhou_full-cycle_2025}, respectively. To focus on the influence of pre-training strategy and dataset composition, we restrict the analysis to a series of MACE models, spanning multiple generations of training approaches \cite{batatia_foundation_2025, batatia_cross_2025}. 

We start with amorphous carbon (Fig.~\ref{fig:figure_3}a--b) which, despite its chemical simplicity, presents a demanding test due to its density-dependent structural diversity. All pre-trained models largely reproduce the local structure (Fig.~\ref{fig:figure_3}b) throughout the density range, recovering the characteristic first peak in the radial distribution function (RDF), near 1.5~\AA{}, and the broader second peak around 2.5~\AA{}, as well as the dominant sp$^2$- and sp$^3$-like features in the angular distribution function (ADF), centred at 120$^\circ$ and 109.5$^\circ$, respectively. 
In contrast, larger discrepancies emerge for ring statistics at low density. At 1.5~g/cm$^3$, \texttt{MACE-MP-0b3} predicts six-membered rings as the dominant motif rather than the expected five-membered rings, resulting in a network topology more characteristic of higher-density $a$-C (and, presumably, of crystalline carbon structures in the training data). The remaining models underestimate the proportion of smaller rings ($n \leq 7$) while overestimating that of larger rings ($n > 15$).

The trends in local structure are reflected in the PCA map of Fig.~\ref{fig:figure_3}a, constructed from SOAP descriptors of the \texttt{C-GAP-17} reference structures. Five distinct density-dependent clusters spanning 1.5--3.5~g/cm$^3$ are formed, and structures generated by each pre-trained MLIP are projected onto this manifold. We assess structural agreement relative to 95\% confidence ellipses constructed from the reference data. Consistent with the ring statistics, the 1.5~g/cm$^3$ pre-trained model structures exhibit the weakest overlap with the reference clusters. \texttt{MACE-MP-0b3} is clearly separated from the remaining models; subsequent MACE models show better structural agreement, but performance does not improve systematically across later model generations (Table S2) -- in contrast to the steady numerical improvements observed in Fig.~\ref{fig:figure_2}.

We next consider $a$-GST (Fig.~\ref{fig:figure_3}c--e), where local structural motifs underpin the characteristic phase-change behaviour \cite{akola_structural_2007, hegedus_microscopic_2008}. Several trends observed for \textit{a}-C are echoed here. Across the GeTe--\ce{Sb2Te3} tie-line, structures generated by pre-trained MLIPs reproduce the short-range order of the reference structures reasonably well, reflected in RDF peak positions and ADF shapes (Fig.~\ref{fig:figure_3}c). The \texttt{MACE-MH-1} structures form an exception -- systematically shorter first-neighbour distances and enhanced first and second RDF peak heights are observed, and the ADF distribution is shifted away from the expected 90$^\circ$ towards larger angles. 

More substantial discrepancies emerge in the network topology. While all models capture the overall increase in ring-size diversity towards the GeTe-rich end of the tie-line, they increasingly favour five-membered rings at the expense of the characteristic ABAB four-membered ring motifs. This bias is most pronounced for \texttt{MACE-MH-1}, whereas \texttt{MACE-MATPES-0} remains in closest agreement with the reference. 
Since four-membered rings play a central role in reversible amorphous--crystalline phase transitions \cite{akola_structural_2007, hegedus_microscopic_2008}, their systematic underprediction suggests that current pre-trained models do not fully capture the structural motifs associated with Ge--Sb--Te phase-change behaviour.

Homopolar bonding motifs (Ge--Ge, Sb--Sb, Ge--Sb, and Te--Te; Fig.~\ref{fig:figure_3}d), which are absent in the corresponding crystalline phases, provide a further probe of structural fidelity. Most pre-trained models reproduce the compositional trend but systematically overestimate the homopolar bond fraction by $\sim$10\%, with \texttt{MACE-MATPES-0} however remaining in close agreement with the reference. Tetrahedrally coordinated Ge atoms (Fig.~\ref{fig:figure_3}e) are likewise overestimated by up to 50\%, consistent with the corresponding overprediction of the short Ge--Ge homopolar bonds that stabilise these environments \cite{deringer_bonding_2014}.
Extended analyses covering the full \textit{a}-C density range, the complete GeTe--\ce{Sb2Te3} tie-line, and \textit{a}-\ce{SiO2} structures generated at multiple quench rates, are provided in the Supplementary Information. 

Together, our results show that current pre-trained MLIPs capture local environments -- often shared between crystalline and amorphous materials and therefore well-represented in existing pre-training datasets -- more reliably than medium-range features that distinguish amorphous networks. Notably, structural fidelity does not correlate clearly with numerical accuracy: despite being the strongest-performing MACE model in the AM26 benchmark (Fig.~\ref{fig:figure_2}), \texttt{MACE-MH-1} exhibits among the poorest agreement with the reference structures. 

\begin{figure*}[t]
    \centering
    \includegraphics[width=\linewidth]{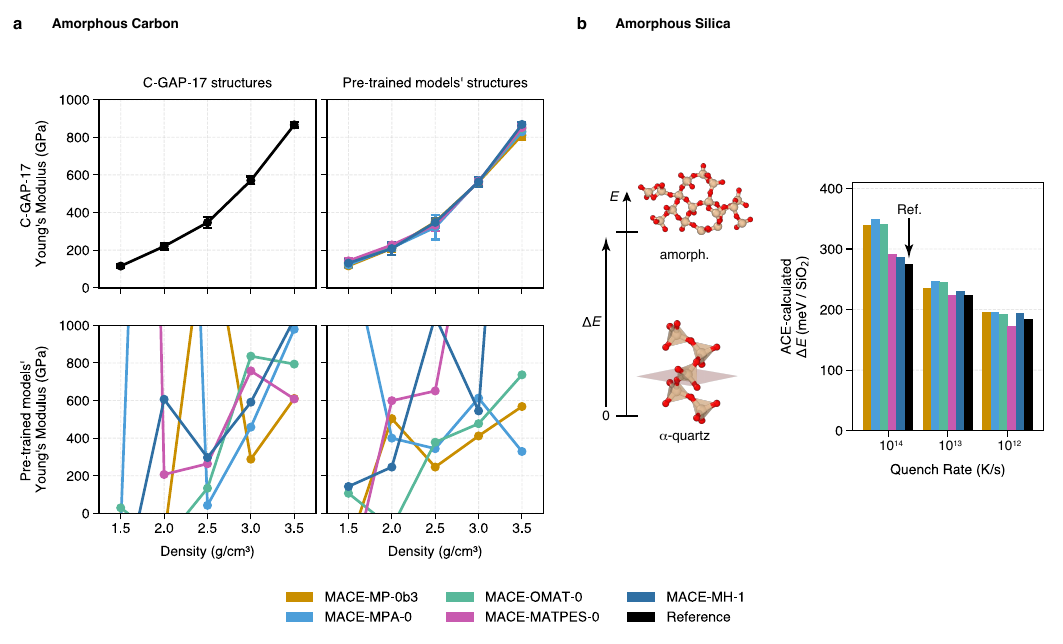}
    \caption{\textbf{Properties of amorphous structures generated by pre-trained MLIPs.}
    (\textbf{a}) Young’s modulus of $a$-C as a function of density, comparing structures generated with the reference \texttt{C-GAP-17} potential and with pre-trained models, and evaluated using either the reference or pre-trained potentials. The mean values over 10 $a$-C structures are displayed, and error bars show one standard deviation. For the bottom row, standard deviations are omitted because their magnitude would obscure the underlying trends, and lines are included as guides to the eye to facilitate comparison with the expected density-dependent trend.
    (\textbf{b}) Excess energies of $a$-\ce{SiO2} structures generated using either the reference or pre-trained models with different quench rates, relative to $\alpha$-quartz. All structures were subsequently relaxed and evaluated using the \texttt{SiO$_x$-ACE-24} potential, and mean values over 10 $a$-\ce{SiO2} structures are shown. A schematic illustrating the excess energy calculation is shown alongside representative structural snapshots of both phases (Si: beige, O: red).} 
    \label{fig:figure_4}
\end{figure*}

\subsection*{Properties}

Using these same generated amorphous structures, we next examine how accurately their underlying potential-energy surfaces are described.
We use the Young's modulus to probe the curvature of the pre-trained potential-energy surfaces (PES) around $a$-C configurations (Fig.~\ref{fig:figure_4}a; see Supplementary Note 2 for computational details). Elastic constants depend on the response of the PES to small applied strains and are therefore particularly sensitive to inaccuracies in its local shape. To establish a baseline, the reference \texttt{C-GAP-17} model was first applied to its own $a$-C structures spanning densities from 1.5--3.5 g/cm$^{3}$ (top left). The resulting Young's moduli reproduce the expected increase in stiffness with density, consistent with the growing sp$^3$ fraction and corresponding approach towards a diamond-like structure, and are in good agreement with experiment (see Ref. \citenum{deringer_machine_2017} and references therein).

To disentangle contributions to the Young's modulus, we use two model choices: one for generating amorphous configurations, and one for relaxing the structures and evaluating their elastic response (either the reference \texttt{C-GAP-17} model, or a pre-trained MLIP). When structures generated by the pre-trained models are evaluated with \texttt{C-GAP-17} (top right), the resulting Young's moduli closely reproduce the reference-on-reference baseline across the density range. This indicates that the generated amorphous configurations occupy physically reasonable regions of the \texttt{C-GAP-17} PES. In contrast, Young's moduli evaluated by the pre-trained models (bottom row) show little correspondence with the reference behaviour, irrespective of whether the underlying structures originate from \texttt{C-GAP-17} or from the pre-trained models themselves. The characteristic increase in stiffness with density is largely absent, with several mean values becoming unphysical and falling below zero. 

These results suggest that errors in Young's modulus predictions from the pre-trained models originate primarily from the curvature of the learned PES around local minima, rather than from the generation of amorphous structures. The contrast with the mostly reasonable structural agreement observed in Fig.~\ref{fig:figure_3}b further suggests that reproducing local structural descriptors does not necessarily imply an accurate description of the underlying PES. Similar failures in elastic-property prediction have been reported for several pre-trained MLIPs in crystalline materials \cite{mannan_evaluating_2025, gao_benchmarking_2026}, suggesting that this limitation is not unique to amorphous systems. Rather, current pre-trained MLIPs appear to be insufficiently constrained around minima more generally.
In this context, incorporating higher-order derivatives into future training protocols, such as stresses, has been proposed to better constrain the PES \cite{mannan_evaluating_2025}.

We next evaluate the excess energies of amorphous silica structures relative to $\alpha$-quartz (Fig.~\ref{fig:figure_4}b). The excess energy provides a complementary probe of the PES by measuring the energetic placement of amorphous configurations relative to the competing crystalline ground state, thereby testing whether the metastable amorphous basin is correctly represented. As above, excess energies are evaluated for structures generated by both the reference \texttt{SiO$_x$-ACE-24} model and pre-trained models. All structures were subsequently relaxed and labelled using the reference model to provide a common energetic baseline.

The reference \texttt{SiO$_x$-ACE-24} model applied to its own structures establishes the expected behaviour (shown with black bars). Excess energies decrease systematically with decreasing quench rate, reflecting the greater degree of structural relaxation achieved during glass formation, and the magnitudes are consistent with experimental estimates for bulk silica glass (see Ref. \citenum{erhard_machine-learned_2022} and references therein). Structures generated by the pre-trained models exhibit the same qualitative quench-rate dependence, and for quench rates of 10$^{12}$ and 10$^{13}$ K/s their excess energies generally remain close to the reference baseline.

More substantial deviations emerge at the fastest quench rate of 10$^{14}$ K/s, where structures generated using \texttt{MACE-MP-0b3}, \texttt{MACE-MPA-0}, and \texttt{MACE-OMAT-0} exhibit excess energies up to $\sim$70 meV/\ce{SiO2} above the reference values. Notably, these energetic offsets persist despite all structures being relaxed to identical convergence criteria on the reference PES. This suggests that under rapid quenching conditions, several pre-trained models sample different regions of configuration space, yielding amorphous configurations that remain energetically less favourable on the reference PES even after relaxation.

\FloatBarrier
\subsection*{Application-level simulations}

\begin{figure*}[t]
    \centering
    \includegraphics[width=\linewidth]{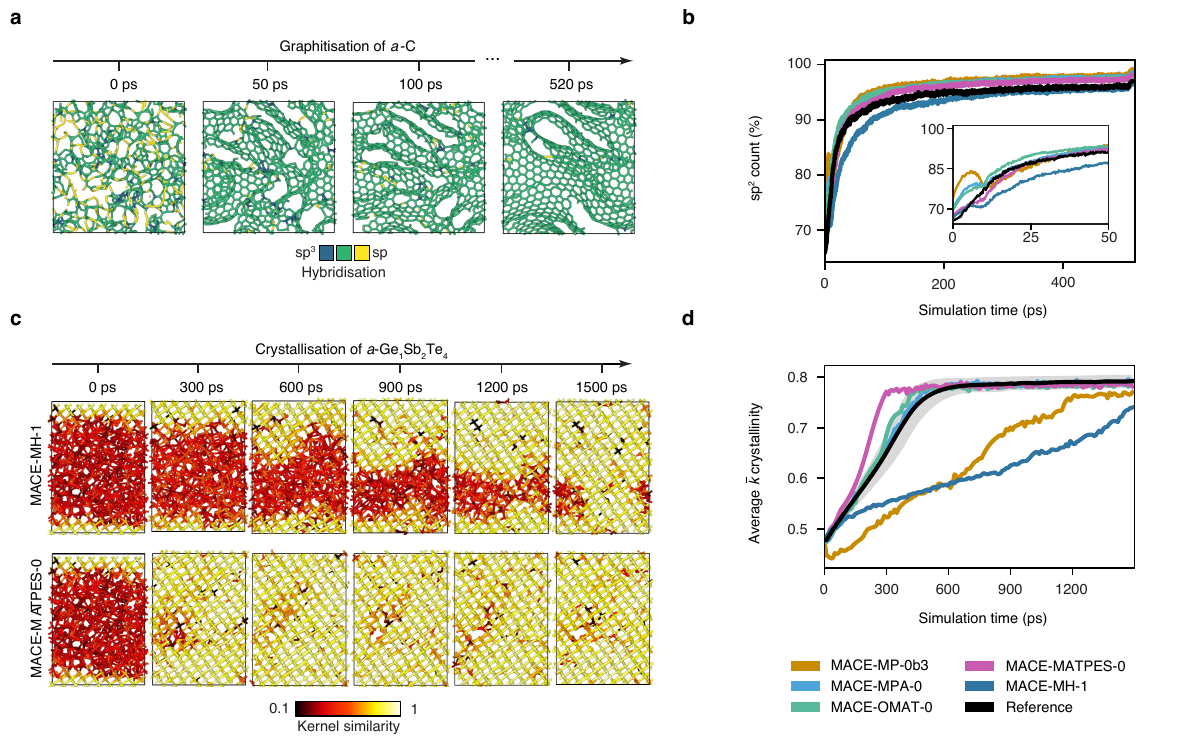}
    \caption{\textbf{Application-level simulations of amorphous materials.}
    (\textbf{a}) Graphitisation of amorphous carbon at 1.5~g/cm$^{3}$ driven by the \texttt{MACE-MP-0b3} potential. A representative slice of the simulation cell is shown. 
    (\textbf{b}) Fraction of sp$^2$ carbon atoms during graphitisation. The inset characterises the first 50~ps of the trajectory.
    (\textbf{c}) Crystallisation of amorphous \ce{Ge1Sb2Te4}, shown for \texttt{MACE-MH-1} (slowest crystallisation) and \texttt{MACE-MATPES-PBE-0} (fastest). Atoms are coloured by a kernel-based crystallinity metric \cite{bartok_representing_2013, xu_unraveling_2022}, with red and yellow indicating amorphous and crystalline environments, respectively.
    (\textbf{d}) Average atomic crystallinity during crystallisation runs. The \texttt{GST-ACE-24} reference is shown as the mean of 100 independent simulations; the grey shaded region indicates the standard deviation. Results in panels (\textbf{b}) and (\textbf{d}) are averaged over three independent simulations for each pre-trained model.}
    \label{fig:figure_5}
\end{figure*}

Finally, we evaluate whether pre-trained MLIPs can reliably sustain and transform amorphous structures over the length- and timescales relevant to practical simulations. 
We first examine the graphitisation of $a$-C (Fig.~\ref{fig:figure_5}a--b). All pre-trained models reproduce the expected behaviour overall, exhibiting rapid graphitisation during the first $\sim$50~ps followed by equilibration at $\sim$95\% sp$^2$ content. Representative snapshots depict the expected formation and growth of graphite-like sheets. Comparable agreement is also observed for compression of $a$-\ce{SiO2} (Supplementary Fig.~S4), indicating that pre-trained MLIPs can reproduce multiple amorphous-phase transformations with good fidelity.

A greater variation emerges for the crystallisation of $a$-GST, corresponding to the SET process (`0 $\rightarrow$ 1') in phase-change memory devices (Fig.~\ref{fig:figure_5}c--d). All pre-trained models capture crystallisation, monitored through an average atomic crystallinity measure (Supplementary Note 2), but at notably different rates. \texttt{MACE-MPA-0} and \texttt{MACE-OMAT-0} crystallise on timescales comparable to the reference potential ($\sim$600~ps), whereas \texttt{MACE-MATPES-0} crystallises in approximately half the time ($\sim$300~ps). In contrast, \texttt{MACE-MP-0b3} and \texttt{MACE-MH-1} do not fully crystallise within the 1.5~ns simulation window. This variation is notable given the common MACE architecture shared by all models, suggesting that pre-training data and training protocol play important roles in determining crystallisation behaviour. In particular, the strong performance of \texttt{MACE-MPA-0} and \texttt{MACE-OMAT-0} is consistent with the broader structural diversity represented in their training data. Conversely, the poor performance of \texttt{MACE-MH-1} in this task, despite its strong numerical accuracy (Fig.~\ref{fig:figure_2}), reinforces a central conclusion of this benchmark: accurate energies and forces alone do not guarantee realistic structure generation, or downstream simulation behaviour.
    
\FloatBarrier
\subsection*{Fine-tuning MLIPs for amorphous materials}

Our benchmarks revealed limitations of zero-shot pre-trained MLIPs across multiple aspects of amorphous materials modelling. A common assumption is that such deficiencies can be readily corrected through domain-specific fine-tuning. We therefore investigate whether fine-tuning can recover amorphous-phase performance, using the early-generation \texttt{MACE-MP-0b3} model as a case study \cite{batatia_foundation_2025}. Of the MACE-family models benchmarked above, it exhibited some of the largest deviations, providing substantial scope for improvement through fine-tuning. Independent fine-tuning datasets for $a$-C, $a$-\ce{SiO2}, and $a$-GST were generated using the same melt-quench procedures employed for AM26, but with newly sampled structures spanning the corresponding density, quench-rate, and composition ranges (Methods).

\begin{figure*}[htbp]
    \centering
    \includegraphics[width=\linewidth]{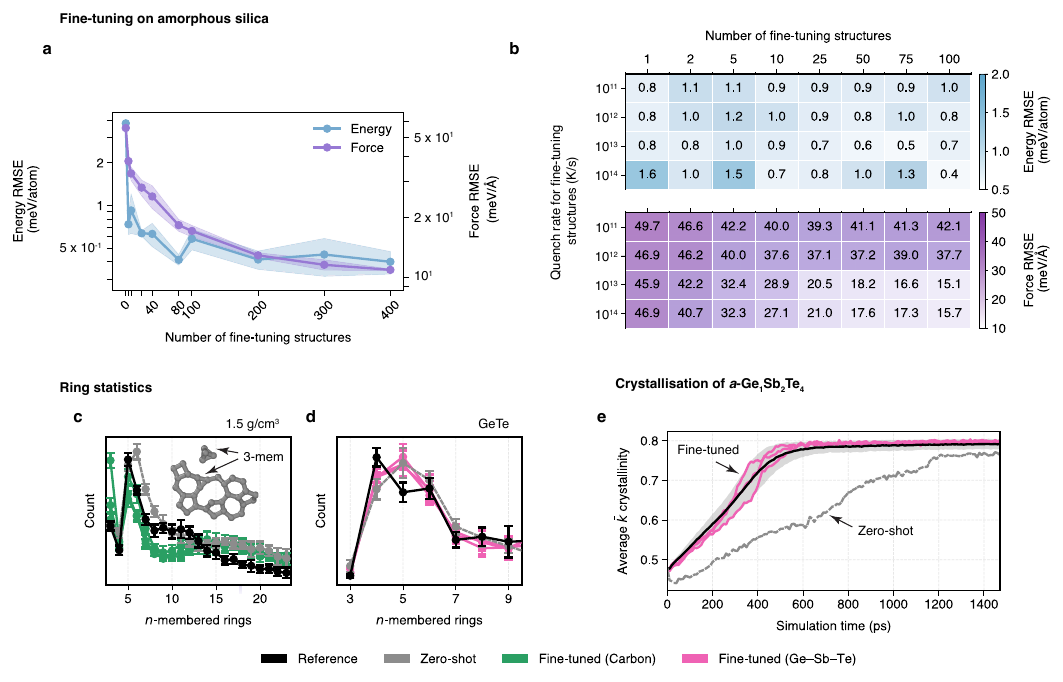}
    \caption{\textbf{Fine-tuning MLIPs in the amorphous domain.}
    We fine-tuned the \texttt{MACE-MP-0b3} model on datasets for the corresponding amorphous target system.
    (\textbf{a}) Energy and force learning curves for amorphous silica. Curves show the mean over three independent fine-tuning repeats; shaded regions indicate the corresponding standard deviation.
    (\textbf{b}) Dependence of $a$-\ce{SiO2} fine-tuning performance on dataset size and quench-rate composition. Rows correspond to fine-tuning datasets constructed from $a$-\ce{SiO2} structures generated at individual quench rates (10$^{11}$--10$^{14}$~K/s), while columns indicate the number of structures used for fine-tuning. Results are averaged over three independent fine-tuning repeats.
    (\textbf{c}) Ring statistics for $a$-C at 1.5~g/cm$^{3}$ and (\textbf{d}) amorphous GeTe structures generated using fine-tuned models, compared against the corresponding zero-shot model.
    (\textbf{e}) Crystallisation trajectories of amorphous \ce{Ge1Sb2Te4}, evaluated through the average atomic crystallinity, for the zero-shot model and fine-tuned variants trained on $a$-GST structures.
    For panels (\textbf{c}--\textbf{e}), three independently fine-tuned models are shown for each system, each trained on the same number of amorphous structures but using different randomly sampled fine-tuning datasets. The number of fine-tuning structures was selected based on saturation of the corresponding learning curves.}
    \label{fig:figure_6}
\end{figure*}

We first examine fine-tuning for amorphous silica. Figure~\ref{fig:figure_6}a shows the corresponding learning curve, where \texttt{MACE-MP-0b3} is fine-tuned on increasing numbers of $a$-\ce{SiO2} structures sampled uniformly across the quench rates in AM26 (10$^{11}$--10$^{14}$~K/s). Each dataset size was evaluated using three independent fine-tuning repeats generated from different data shuffles, with reported values corresponding to the mean across repeats. Fine-tuning rapidly improves both energy and force predictions on the AM26 $a$-\ce{SiO2} subset (Fig.~\ref{fig:figure_6}a). The largest improvements occur for training sets containing fewer than $\sim$100 structures, beyond which gains become progressively smaller. With only four $a$-\ce{SiO2} structures, the energy error is reduced by more than a factor of five relative to the zero-shot \texttt{MACE-MP-0b3}, while the force error is nearly halved. Substantial improvements in amorphous-domain numerical accuracy can therefore be achieved with only modest amounts of domain-specific training data.

The composition of the fine-tuning dataset strongly affects performance (Fig.~\ref{fig:figure_6}b). Energy errors decrease substantially from the zero-shot baseline ($\sim$4~meV/atom) after fine-tuning on even a single amorphous structure produced with any quench rate, with no systematic further improvement as dataset size or composition is varied. In contrast, force predictions remain strongly dependent on both factors. Fine-tuning datasets constructed from rapidly quenched structures (10$^{13}$--10$^{14}$~K/s) consistently outperform those derived from slower quenches (10$^{11}$--10$^{12}$~K/s), reducing force errors by almost a factor of two once the dataset contains $\geq$25 structures. This trend is somewhat counterintuitive: although slower-quenched structures are thermodynamically more favourable (see Fig.~\ref{fig:figure_4}b), rapid quenching likely provides more informative training data by sampling a broader range of local environments and forces. 

We next examine whether improved numerical accuracy translates into improved amorphous structures. Similar learning curves were obtained for $a$-C and GST (Supplementary Fig.~S5). Models from the plateau regions of these curves (100--200 structures) were selected for subsequent evaluation. For low-density $a$-C (1.5 g/cm$^{3}$), the zero-shot \texttt{MACE-MP-0b3} model substantially mispredicted the ring distribution, overrepresenting six-membered rings relative to five-membered ones, and predicting an excess of large rings ($n \geq 15$; Fig.~\ref{fig:figure_6}c). Fine-tuning improves the five-to-six-membered ring ratio in all three training repeats. However, these improvements are accompanied by new deviations, including the underprediction of medium-sized rings, while the overprediction of larger rings remains unresolved. Notably, one fine-tuned model produces a pronounced excess of three-membered rings, arising from unphysical clusters dispersed throughout the amorphous network (shown in the inset of Fig.~\ref{fig:figure_6}c). This behaviour is particularly striking since all three fine-tuned models exhibit comparable numerical accuracy on the AM26 $a$-C benchmark (Supplementary Fig.~S6), demonstrating that similar energy and force errors can nevertheless yield qualitatively different amorphous networks.
In contrast to the partial improvements observed for $a$-C, fine-tuning has little effect on the predicted network topology of $a$-GST. The overprediction of five- relative to four-membered rings observed in the zero-shot model persists after fine-tuning (Fig.~\ref{fig:figure_6}d).

Finally, we revisit the GST crystallisation benchmark of Fig.~\ref{fig:figure_5}c--d using fine-tuned MLIPs. Whereas the zero-shot \texttt{MACE-MP-0b3} model failed to complete crystallisation within the 1.5~ns simulation window, all three fine-tuned models reproduce the crystallisation behaviour of the reference ACE potential ($\sim$600~ps). This result is notable for two reasons. First, this behaviour is recovered using only 112 fine-tuning structures, despite specialised GST MLIPs, such as \texttt{GST-GAP-22} \cite{zhou_device-scale_2023} and \texttt{GST-ACE-24} \cite{zhou_full-cycle_2025}, being trained on substantially more data (2,692 and 4,636 structures, respectively). Second, the restoration of crystallisation behaviour occurs despite the persistence of deviations in the higher-order structural features of $a$-GST (Fig.~\ref{fig:figure_6}d), indicating that successful reproduction of application-level behaviour does not necessarily require recovery of all structural descriptors.

Overall, these results show that fine-tuning provides a powerful but incomplete strategy for adapting pre-trained MLIPs to amorphous materials. While modest amounts of domain-specific data can substantially improve numerical accuracy and recover some application-level behaviour, they do not necessarily improve the generated amorphous structures and may introduce new structural artefacts. This suggests that some limitations of current pre-trained models cannot be straightforwardly remedied through domain adaptation alone, and that alternative approaches, including improved representation of disordered environments during pre-training, may be required.

\section*{Discussion}
Achieving truly universal behaviour remains a central challenge in MLIP development. Our results have revealed substantial limitations in the transferability of many current pre-trained MLIPs to amorphous materials, and they show that fine-tuning does not provide a consistent solution across materials systems and benchmarking tasks. Progress in modelling crystalline materials has been driven by the combination of large-scale datasets \cite{deng_chgnet_2023, schmidt_improving_2024, barros-luque_open_2026, kaplan_foundational_2025, levine_open_2026, gharakhanyan_open_2026} and systematic benchmarking \cite{dunn_benchmarking_2020, riebesell_framework_2025, choudhary_jarvis-leaderboard_2024}; yet, amorphous materials have so far largely been absent from these initiatives. By providing a common benchmark for amorphous materials, AM26 allows future models to be systematically evaluated and improved in this domain.

Our results have implications for the design of next-generation pre-training datasets. Scale alone does not guarantee transferability to the amorphous state -- instead, future datasets require broader coverage of distinct regions of configuration space, including representative amorphous configurations rather than relying primarily on disordered structures derived from crystalline parents through perturbation or MD \cite{barros-luque_open_2026}. Achieving such a broad coverage will likely require the same level of systematic data generation that has previously been devoted to crystalline materials \cite{zheng_ab_2024, liu_medium-range_2026}. The observed challenges in elastic-property prediction (Fig.~\ref{fig:figure_4}a) further indicate that higher-order quantities, such as stresses, should be incorporated as training labels, and they emphasise the importance of physically-based evaluations \cite{morrow_how_2023}. 

We conclude by noting that many real-world materials occupy a continuum of structural disorder, with partially ordered and polycrystalline structures that lie between the idealised limits of fully crystalline and fully amorphous phases \cite{rosset_signatures_2025, pasca_machine-learning-driven_2025}. As such, genuinely universal MLIPs will require the ability to describe both order and disorder within a single transferable framework. Achieving this would enable increasingly powerful tools for the modelling -- and ultimately design -- of amorphous functional materials \cite{liu_amorphous_2024}, while substantially broadening the scope of foundation-model approaches across materials science.

\clearpage
\setstretch{1.1}
\section*{Methods}
\subsection*{The AM26 dataset}
AM26 is a dataset of amorphous structures designed to benchmark MLIPs in this domain (Table \ref{tab:a26-dataset}). Amorphous Si and Ge--Sb--Te structures were sourced from Refs.~\citenum{rosset_signatures_2025} and \citenum{zhou_device-scale_2023}, respectively, and were subsequently filtered by structure origin and size, with densely sampled regions subsampled to reduce redundancy (Supplementary Note 1), before being re-labelled with consistent DFT settings. For amorphous carbon, silica, and Li--P--S, new structures were generated using melt–quench protocols with validated specialist potentials \cite{deringer_machine_2017, erhard_modelling_2024, fragapane_lips_2026} (Supplementary Note 2), as existing datasets were primarily developed via iterative MLIP training and thus contained highly distorted configurations. This combination yields a structurally representative benchmark spanning diverse classes of amorphous materials.

\begin{table}[hb]
    \caption{Composition of the AM26 dataset. Reported for each amorphous system are the number of structures ($N_\mathrm{cells}$), the total number of atoms ($N_\mathrm{atoms}$), and a description of each dataset.}
    \centering
    \begin{threeparttable}
        \begin{tabular}{lccp{10.5cm}}
            \hline
            \textbf{Dataset} & 
            \textbf{\textit{N}$_{\!\textbf{cells}}$} & 
            \textbf{\textit{N}$_{\!\textbf{atoms}}$} & 
            \textbf{Description} \\
            \hline
            \textit{a}-C           & 100 & 21,600 & Melt-quenched structures (density range: 1.5--3.5 g/cm$^{3}$), generated using the \texttt{C-GAP-17} model \cite{deringer_machine_2017}.  \\ 
            \textit{a}-Si          & 324 & 45,056 & Filtered subset of the a-Si-24 dataset from Ref.~\citenum{rosset_signatures_2025}.  \\ 
            \textit{a}-\ce{SiO2}   & 80 & 24,000 & Melt-quenched structures (quench rates: 10$^{11}$, 10$^{12}$, 10$^{13}$, 10$^{14}$ K/s) generated using the \texttt{SiO$_{x}$-ACE-24} model \cite{erhard_modelling_2024}. \\ 
            \textit{a}-GST         & 290 & 65,980 & Combination of AIMD configurations from GST-GAP-22 and the external validation set of Ref.~\citenum{zhou_device-scale_2023}, spanning compositions along the GeTe--Sb$_2$Te$_3$ tie-line. \\ 
            \textit{a}-LiPS        & 140 & 41,360 & Melt-quenched structures generated using a model from Ref.~\citenum{fragapane_lips_2026}, spanning compositions along the Li$_2$S--P$_2$S$_5$ tie-line. \\ 
            \hline
            \textbf{Total}         & 934 & 197,996  &  \\ 
            \hline
        \end{tabular}
    \end{threeparttable}
    \label{tab:a26-dataset}
\end{table}

\subsection*{Reference computations}
Density-functional theory (DFT) labels for the AM26 and fine-tuning datasets were generated following the default settings of the Materials Project \cite{jain_commentary_2013} using VASP 6.4.3. \cite{kresse_efficient_1996, kresse_efficiency_1996, kresse_ab_1994, kresse_ab_1995}. We used the projector augmented-wave method \cite{blochl_projector_1994, kresse_ultrasoft_1999}, the Perdew--Burke--Ernzerhof (PBE) exchange--correlation functional \cite{perdew_generalized_1996}, and the VASP PBE 5.4 pseudopotential set. This choice is consistent with several recent large-scale datasets, including OMat24 \cite{barros-luque_open_2026}, and differs from earlier datasets such as MPTrj \cite{deng_chgnet_2023}, which used older VASP pseudopotentials \cite{horton_accelerated_2025}. A plane-wave energy cutoff of 520 eV was applied, and total energies were converged to within $5 \times 10^{-5}$ eV/atom. Further details are provided in the Supplementary Information.

\subsection*{Benchmarking framework and evaluation metrics}
Numerical accuracy on the AM26 dataset is assessed using the relative root-mean-square error (RelRMSE), defined for a each system as

\begin{equation}
\mathrm{RelRMSE} = 100 \times
\frac{\sqrt{\frac{1}{N}\sum_{i=1}^{N} \left(y_i - \hat{y}_i\right)^2}}{\sigma_y},
\end{equation}

where \(y_i\) and \(\hat{y}_i\) denote the reference and predicted values, respectively, \(N\) is the number of structures, and \(\sigma_y\) is the standard deviation of the reference labels. RelRMSE values are computed independently for each system, and overall model performance is reported as the mean RelRMSE across all five systems.

For structure generation and application tests, model performance is evaluated against established specialist MLIPs: C-GAP-17 for amorphous carbon \cite{deringer_machine_2017}, SiO$_x$-ACE-24 potential for amorphous silica \cite{erhard_modelling_2024}, and GST-ACE-24 potential for amorphous Ge–Sb–Te \cite{zhou_full-cycle_2025}. 

Detailed definitions and implementation details for the structural and property-based metrics used to evaluate generated structures and application-level tests are provided in the Supplementary Information.

\subsection*{Pre-trained MLIPs}
We evaluated 41 pre-trained MLIPs spanning ten architectural families; in alphabetical order: GRACE \cite{lysogorskiy_graph_2026}, MACE \cite{batatia_foundation_2025, batatia_cross_2025}, MatRIS \cite{zhou_matris_2026}, MatterSim \cite{yang_mattersim_2024}, NequIP \cite{tan_high-performance_2025}, Orb \cite{neumann_orb_2024, rhodes_orb-v3_2025}, PET-OAM \cite{mazitov_pet-mad_2025}, SevenNet \cite{park_scalable_2024}, DeePMD \cite{zhang_graph_2026}, and UMA \cite{wood_uma_2026}. Most of these models are trained on large-scale, general-purpose pre-training datasets generated from DFT computations with broadly consistent settings, including MPTrj \cite{deng_chgnet_2023}, sAlex \cite{schmidt_improving_2024}, OMat24 \cite{barros-luque_open_2026}, MATPES \cite{kaplan_foundational_2025}, and OpenLAM \cite{peng_openlam_2025}. The MatterSim family is an exception, having been trained on a closed dataset generated via active learning, with configurations iteratively selected to span a wide range of temperatures, pressures, and chemistries \cite{yang_mattersim_2024}.

The models can be broadly categorised according to their training strategy. Many are trained directly on one or more of the above datasets, in some cases followed by fine-tuning. In contrast, some employ multi-head architectures, in which separate output heads are specialised to data at different levels of theory, enabling cross-domain knowledge transfer while retaining regime-specific accuracy. These include \texttt{MACE-MH-1} \cite{batatia_cross_2025} and the SevenNet multi-fidelity variants (\texttt{SevenNet-omni} and \texttt{SevenNet-MF}) \cite{park_scalable_2024}. UMA models represent a third category, using a unified multi-task architecture trained jointly across multiple DFT datasets, with task identity (e.g., level of theory, charge, and spin state) provided as global conditioning information and labels normalised to enable stable joint training \cite{wood_uma_2026}.

\subsection*{Fine-tuning}
Pre-trained MLIPs can be adapted to specific chemical and configurational domains through fine-tuning, enabling improved accuracy in targeted regimes, correction of systematic biases such as PES softening in off-equilibrium domains \cite{deng_systematic_2025}, or adaptation to different levels of electronic structure theory \cite{smith_approaching_2019}. In this work, we employ a direct (``naive'') fine-tuning protocol, in which all model weights are updated through further training on the fine-tuning dataset. Recent work has shown that such approaches remain highly effective for domain-specific adaptation tasks, while more involved strategies such as multi-head replay fine-tuning \cite{batatia_foundation_2025} (\url{https://github.com/ACEsuit/mace}) and frozen-transfer methods \cite{radova_fine-tuning_2025} primarily benefit retention of performance outside the fine-tuning domain \cite{tompa_fine-tuning_2026}.

Fine-tuning of the \texttt{MACE-MP-0b3} model was performed using the \texttt{graph-pes} package \cite{gardner_graph-pes_2024}. We used the AdamW optimiser \cite{loshchilov_decoupled_2019} with a learning rate of $10^{-4}$ and a batch size of 5. The loss function combined energy and force terms in a 1:10 ratio, consistent with the original training procedure in Ref.~\citenum{batatia_foundation_2025}. Early stopping was applied based on the validation loss, with training terminated after 50 epochs without improvement.

Learning curves were constructed to assess data efficiency, and additional analyses were performed to probe the effect of fine-tuning dataset composition. Detailed descriptions of training procedures, fine-tuning dataset construction, and sampling protocols are provided in Supplementary Note 2.

\section*{Data availability}
Data supporting this work are available at \url{https://github.com/vldgroup/AM26}.

\section*{Acknowledgements}
We thank Louise A. M. Rosset and Yuxing Zhou for helpful discussions. This work was supported by UK Research and Innovation [grant number EP/X016188/1]. We are grateful for computational support from the UK national high performance computing service, ARCHER2, for which access was obtained via the UKCP consortium and funded by EPSRC grant ref EP/X035891/1 (see also Ref.~\citenum{beckett_archer2_2024}).

\end{document}


\title{
\fontsize{20}{26}\selectfont
\textbf{Supplementary Information for}\\
`Amorphous materials as a frontier challenge for universal interatomic potentials'
}

\author[1]{Natascia L. Fragapane}
\author[1]{Volker L. Deringer\thanks{volker.deringer@chem.ox.ac.uk}}

\affil[1]{Inorganic Chemistry Laboratory, Department of Chemistry, University of Oxford,\protect\\ Oxford, UK}

\date{}

\maketitle

\setstretch{1.1}

\clearpage
\section*{Supplementary Note 1: \normalfont{Components of the AM26 dataset}}
Details of the construction of each subset of AM26 are provided below.

\textbf{Carbon.}
Structures for the \textit{a}-C subsection of AM26 were generated via melt–quench simulations driven by the \texttt{C-GAP-17} potential \cite{deringer_machine_2017}, following the protocol described in Supplementary Note 2. At each density value of 1.5, 2.0, 2.5, 3.0, and 3.5 g/cm$^3$, 20 independent \textit{a}-C structures containing 216 atoms were produced using distinct random seeds for velocity initialisation.

\textbf{Silicon.}
We construct the \textit{a}-Si subset of AM26 by filtering the existing high-quality ``a-Si-24'' dataset provided by Rosset et al.\ in Ref.~\citenum{rosset_signatures_2025}. The original dataset contains 3,069 \textit{a}-Si structures, each corresponding to the final snapshot of an independent melt–quench trajectory. It spans quench rates of 10$^{10}$–10$^{13}$ K/s, and densities between 2.1 and 2.5 g/cm$^{3}$ in increments of 0.002 g/cm$^{3}$, covering a range of configurations from amorphous to partly crystalline (see Ref.~\citenum{rosset_signatures_2025} and references therein).

To reduce computational cost while retaining representative structural diversity, a two-stage filtering procedure was applied in the present work. First, structures were filtered by system size. The original a-Si-24 dataset contains 801 structures with 64 atoms, 782 with 216 atoms, 797 with 512 atoms, and 689 with 1,000 atoms. Structures containing $\geq$ 512 atoms were excluded, leaving only the 64- and 216-atom configurations.
Second, the remaining structures were filtered by density, retaining only those whose densities correspond to discrete values spaced by 0.01 g/cm$^{3}$ within the 2.1--2.5 g/cm$^{3}$ range (i.e., densities of 2.10, 2.11, ... , 2.50 g/cm$^{3}$). This selection reduces redundancy while preserving uniform coverage of the density range. This procedure yielded 164 structures with 64 atoms and 160 structures with 216 atoms, which together define the \textit{a}-Si component of AM26.

\textbf{Silica.}
Structures for the \textit{a}-\ce{SiO2} subset of AM26 were generated via melt–quench simulations driven by the \texttt{SiO$_{x}$-ACE-24} potential \cite{erhard_modelling_2024}, following the protocol described in Supplementary Note 2. 20 independent amorphous silica structures containing 300 atoms were produced using distinct random seeds for velocity initialisation. The starting structures had a density of 2.2 g/cm$^{3}$, corresponding to the experimental density of amorphous silica \cite{mazurin_silica_2010}.

\textbf{Ge--Sb--Te.}
The \textit{a}-GST subset of AM26 is constructed by combining two datasets from previous work by Zhou et al.\ (Ref.~\citenum{zhou_device-scale_2023}): (i) the ``External Validation'' set used for numerical validation, and (ii) the AIMD configurations included in the GST-GAP-22 training dataset. Both datasets comprise snapshots extracted from AIMD melt--quench trajectories spanning compositions along the \ce{GeTe}--\ce{Sb2Te3} tie-line (\ce{Sb2Te3}, \ce{Ge1Sb4Te7}, \ce{Ge2Sb2Te5}, \ce{Ge3Sb2Te6}, \ce{Ge4Sb2Te7}, \ce{Ge8Sb2Te11}, and \ce{GeTe}). The ``External Validation'' set contains 80 disordered configurations that were not included in the original GST-GAP-22 training data. Full details of the AIMD simulations can be found in the original work \cite{zhou_device-scale_2023}. 

\textbf{Li--P--S.}
Structures for the \textit{a}-\ce{LiPS} subset of AM26 were generated via melt–quench simulations driven by the MACE potential with a cut-off of 6~\AA{} reported in Ref.~\citenum{fragapane_lips_2026}. Twenty independent amorphous Li–P–S structures containing approximately 200–300 atoms were produced for each composition along the \ce{Li2S}–\ce{P2S5} tie-line (\ce{Li2S}, \ce{Li7PS6}, \ce{Li3PS4}, \ce{Li7P3S11}, \ce{Li4P2S7}, \ce{Li2P2S6}, and \ce{P2S5}) using distinct random seeds for velocity initialisation. Initial configurations were constructed from the corresponding crystalline phases expanded into supercells of approximately 300 atoms: \ce{Li2S} (ICSD 60432), \ce{Li7PS6} (mp-1211324), $\gamma$-\ce{Li3PS4} (ICSD 180318), \ce{Li7P3S11} (ICSD 157654), \ce{Li4P2S7} (Ref.~\citenum{holzwarth_computer_2011}), \ce{Li2P2S6} (ICSD 253894), and \ce{P2S5} (ICSD 409061).
The melt–quench protocol comprised an initial 25 ps annealing run at 300 K, heating to 1500 K over 50 ps, a 25 ps hold at 1500 K, cooling back to 300 K over 50~ps, and a final 25 ps anneal at 300 K. The final snapshot of each trajectory was taken as the amorphous structure. All stages were performed within the NVT ensemble using a timestep of 1 fs and a thermostat damping parameter of \(t_{\text{damp}}^{(T)} = 100\)~fs.

\clearpage

\section*{Supplementary Note 2: \normalfont{Computational details}}
\subsection*{DFT computations}
All density-functional theory (DFT) computations were performed using the Vienna Ab initio Simulation Package (VASP, version 6.5.1) \cite{kresse_ab_1994, kresse_efficiency_1996, kresse_efficient_1996}. Input files were generated using the \texttt{MPStaticSet} workflow implemented in \texttt{pymatgen} \cite{ong_python_2013}, following the standard Materials Project protocols (see Ref.~ \citenum{horton_accelerated_2025} and \url{https://github.com/materialsproject/pymatgen/}). The projector augmented wave (PAW) method \cite{blochl_projector_1994, kresse_ultrasoft_1999} was employed with the Perdew--Burke--Ernzerhof (PBE) exchange-correlation functional \cite{perdew_generalized_1996}, with version 5.4 pseudopotentials. This is consistent with newer datasets, including OMat24 \cite{barros-luque_open_2026}, but differs from earlier datasets such as MPTrj \cite{deng_chgnet_2023}, which used older pseudopotential versions for several elements. 

A plane-wave energy cutoff of 520 eV was used throughout. Brillouin zone integrations were performed using $\Gamma$-centred Monkhorst–Pack grids with a $k$-point density consistent with Materials Project standards, as implemented in \texttt{MPStaticSet}. Electronic self-consistency was achieved with an energy convergence criterion of $5 \times 10^{-5}$ eV per atom. Smearing was handled using the default Materials Project settings. These settings ensure consistency across all labelled datasets and are broadly consistent with the pre-training datasets used by the MLIPs evaluated in this work.

\subsection*{MD simulations}

All molecular-dynamics (MD) simulations were performed using LAMMPS \cite{thompson_lammps_2022}. Established melt--quench protocols were employed for each material, following the procedures outlined in Ref.~\citenum{deringer_machine_2017} ($a$-C), Ref.~\citenum{erhard_machine-learned_2022} ($a$-\ce{SiO2}), and Ref.~\citenum{zhou_device-scale_2023} ($a$-GST), respectively. Domain-specific simulation tasks for selected materials were used to further assess model performance for selected materials. Details of both types of MD simulations are given in the following.

\subsubsection*{Melt--quenching for initial structure generation}

Amorphous carbon structures were generated following Ref.~\citenum{deringer_machine_2017}. Initial configurations were constructed by placing carbon atoms on a simple-cubic lattice at the target density (1.5 -- 3.5 g/cm$^3$). The systems were first held at 9000~K for 3~ps to remove memory of the initial configuration. The temperature was then held at 5000~K for a further 3~ps, before being quenched to 300~K over 0.5~ps with an exponentially decaying temperature profile. Finally, the structures were annealed at 300~K for an additional 3~ps. All simulations were performed using a Langevin thermostat \cite{bussi_canonical_2007}, with time integration carried out in the NVE ensemble, ensuring constant density throughout the trajectory. A timestep of 1~fs and thermostat damping parameter \(t_{\text{damp}}^{(T)} = 100\)~fs were used.

Amorphous \ce{SiO2} structures were generated following Ref.~\citenum{erhard_machine-learned_2022}. Initial configurations were constructed by randomly placing Si and O atoms, with the appropriate stoichiometric ratio, at the target density of 2.2 g/cm$^3$. The systems were first randomised at 6000~K for 10~ps within the NVT ensemble. The temperature was then held at 4000 K for 100 ps within the NPT ensemble at zero external pressure to generate the melt. The melt was subsequently quenched to 300~K at rates of $10^{11}$, $10^{12}$, $10^{13}$, or $10^{14}$ K/s, as indicated, and then held at 300 K for a further 10~ps. A Nosé--Hoover thermostat \cite{nose_molecular_1984, hoover_canonical_1985} was used for NVT MD, with a Nosé--Hoover barostat \cite{martyna_constant_1994} applied during NPT ensembles. A timestep of 1~fs was employed throughout, with thermostat and barostat damping parameters of \(t_{\text{damp}}^{(T)} = 100\)~fs and \(t_{\text{damp}}^{(p)} = 1000\)~fs, respectively.

Amorphous Ge--Sb--Te structures were generated following Ref.~\citenum{zhou_device-scale_2023}. Initial configurations were constructed by randomly placing Ge, Sb, and Te atoms in the desired stoichiometric ratio at the theoretical density of the corresponding amorphous structure, as obtained at the PBE level. The systems were first randomised at 3000~K for 20~ps, followed by a quench to 1000~K at a rate of $10^{14}$~K/s, after which the system was held at 1000~K for 50~ps. The melt was then quenched to 300~K at a rate of 2.5 $\times$ $10^{13}$~K/s, and held at 300~K for a further 50~ps. Simulations were performed using a Langevin thermostat \cite{bussi_canonical_2007}, with time integration carried out in the NVE ensemble. A timestep of 2~fs and a thermostat damping parameter of \(t_{\text{damp}}^{(T)} = 40\)~fs were used throughout. 

\subsubsection*{Domain-specific simulation tasks}

Carbon graphitisation simulations were performed starting from 4096-atom \textit{a}-C configurations at 1.5 g/cm$^3$, generated using the melt–quench protocol described above with the \texttt{C-GAP-17} potential \cite{deringer_machine_2017}. Each system was heated from 300~K to 3000~K over 10~ps, annealed at 3000~K for 500~ps to facilitate graphitisation, and subsequently quenched back to 300~K over 10~ps. Simulations were conducted in the NVT ensemble using a Nosé–Hoover thermostat \cite{nose_molecular_1984, hoover_canonical_1985}, with a timestep of 1~fs and a thermostat damping parameter of \(t_{\text{damp}}^{(T)} = 100\)~fs. For each potential, three independent runs were performed using different random seeds for velocity initialisation; the results reported correspond to the averages over these repeats.

For simulating the compression behaviour of \textit{a}-\ce{SiO2}, the protocol of Refs.~\citenum{erhard_machine-learned_2022} was followed. The simulation started from a 5184-atom \textit{a}-\ce{SiO2} structure generated via the melt--quench procedure described above, using the \texttt{SiO$_{x}$-ACE-24} potential with a quench rate of 10$^{13}$ K/s. The system was annealed for 20~ps at 300~K and 0~GPa, followed by isostatic compression to 70~GPa over 20~ps. The NPT ensemble was used with a Nosé--Hoover thermostat \cite{nose_molecular_1984, hoover_canonical_1985} and a Nosé--Hoover barostat \cite{martyna_constant_1994}. A timestep of 1~fs was employed throughout, with thermostat and barostat damping parameters of \(t_{\text{damp}}^{(T)} = 100\)~fs and \(t_{\text{damp}}^{(p)} = 1000\)~fs, respectively.

The crystallisation of \ce{Ge1Sb2Te4}, corresponding to the `SET' process in phase-change memory devices (`0 $\rightarrow$ 1'), was simulated following the protocol of Ref.~\citenum{zhou_device-scale_2023}. A 1008-atom \textit{a}-\ce{Ge1Sb2Te4} structure was annealed at 600~K for 1.5~ns using a Langevin thermostat \cite{bussi_canonical_2007}, while time integration was performed in the NVE ensemble. A timestep of 2~fs and a thermostat damping parameter of \(t_{\text{damp}}^{(T)} = 100\)~fs were used throughout. 

\subsection*{Structural analysis}
RDFs and ADFs were determined using OVITO \cite{stukowski_visualization_2009}. Coordination numbers were calculated with ASE \cite{hjorth_larsen_atomic_2017} or OVITO \cite{stukowski_visualization_2009} for larger structures. The bond-length cut-offs used for ADFs and coordination numbers were 1.85~\AA{} for $a$-C, 2.0~\AA{} for $a$-\ce{SiO2}, and 3.2~\AA{} for GST. Ring counts were obtained using Franzblau's shortest-path algorithm \cite{franzblau_computation_1991} as implemented in \texttt{matscipy} \cite{grigorev_matscipy_2024}. For $a$-C structures, RDFs and ADFs were evaluated with 512-atom structures, and rings with larger 4096-atom structures. For $a$-GST, RDFs, ADFs, and rings were evaluated with $\sim$500-atom structures (varying with stoichiometry). 

The structural similarity of $a$-C structures and the crystallisation simulation of \ce{Ge1Sb2Te4} were both analysed using Smooth Overlap of Atomic Positions (SOAP) descriptors \cite{bartok_representing_2013}. SOAP descriptors represent local atomic environments through an expansion of the neighbour density into a basis of radial functions and spherical harmonics within a finite radial cut-off, with Gaussian smearing applied to atomic positions. SOAP provides a rotationally invariant representation of local atomic environments that can be compared via similarity measures.

The result of such an analysis for $a$-C structures is visualised in Fig.~3a in the main text. For this case, SOAP hyperparameters were chosen to be consistent with those employed in the training of the \texttt{C-GAP-17} potential ($r_{\mathrm{cut}}$ = 3.7~\AA{}, $n_{\mathrm{max}}$ = 8, $l_{\mathrm{max}}$ = 8, $\sigma$ = 0.5~\AA{}) \cite{deringer_machine_2017}. For each structure, a global descriptor was constructed by averaging local SOAP vectors (`outer' averaging) over all atoms. A principal component analysis (PCA) embedding \cite{lovric_principal_2011, pedregosa_scikit-learn_2011} was obtained by fitting a two-dimensional manifold to the global SOAP descriptors of 100 reference $a$-C structures at each density generated with \texttt{C-GAP-17} (500 structures in total). SOAP descriptors for 100 structures generated by each pre-trained model at each density were subsequently projected onto this PCA space, enabling direct comparison. The distribution of reference structures per density was represented using 95\% confidence ellipses in the PCA space, constructed from the covariance of the projected reference descriptors within each density cluster. The resulting projection provides a low-dimensional representation of the SOAP feature space, in which the proximity of pre-trained model structures to the reference distributions reflects their structural similarity.

To quantify the per-atom `crystallinity' throughout the crystallisation simulation shown in Fig.~5c of the main text, SOAP descriptors were again employed (hyperparameters consistent with those used in the same analysis in Refs.~\citenum{xu_unraveling_2022} and \citenum{zhou_device-scale_2023}: $r_{\mathrm{cut}}$ = 9.0~\AA{}, $n_{\mathrm{max}}$ = 16, $l_{\mathrm{max}}$ = 16, $\sigma$ = 0.3~\AA{}). Each atomic environment was compared to those in three reference structures of idealised rocksalt-like \ce{Ge1Sb2Te4}, in which Ge and Sb atoms, as well as vacancies, are randomly distributed over the cation sublattice. Atoms are grouped by chemical identity, with Ge/Sb treated as cations and Te as anions, and comparisons are performed only within the same group. SOAP similarities between each atom in the simulation and all corresponding environments in the reference set are computed and averaged to obtain the `$\bar{k}$ crystallinity' measure, where $\bar{k}$ = 1 indicates ideal crystallinity, as in Refs.~\citenum{xu_unraveling_2022} and \citenum{zhou_device-scale_2023}. For each MD snapshot, the $\bar{k}$-crystallinity is further averaged over all atoms in the simulation cell, yielding a time-resolved crystallinity measure that is plotted as a function of simulation time in Fig.~5d.

The counts of tetrahedrally coordinated Ge atoms and homopolar bonds were calculated as averages over the last 40~ps of 10 independent melt--quench trajectories. The mean value was plotted, and the standard deviation indicated with error bars. Tetrahedral Ge coordination is defined using a bond-order parameter introduced in Ref.~\citenum{chau_new_1998} and previously applied to GST systems in Refs.~\citenum{caravati_coexistence_2007} and \citenum{zhou_device-scale_2023}. Homopolar bonds are defined as bonds between two cation-like or two anion-like atoms (Ge--Ge, Sb--Sb, Ge--Sb, Te--Te).

The Young's modulus of $a$-C structures was calculated as an average over ten 512-atom configurations generated via the melt--quench procedure described above. Each structure was further quenched to near 0~K and relaxed using a conjugate-gradient algorithm with fixed cell vectors to preserve the density. The $6 \times 6$ elastic-constant matrix, \(\mathbf{C}\), was then computed and inverted to obtain the compliance matrix, \(\mathbf{S}\), from which the Young's modulus, \textit{E}, was calculated by averaging over three spatial directions:

\begin{equation}
E = \frac{1}{3} \left[ \frac{1}{S_{11}} + \frac{1}{S_{22}} + \frac{1}{S_{33}} \right]
\end{equation}

The reported value was subsequently averaged over the ten independent structures. 

To analyse the origin of discrepancies between models, this protocol was systematically decomposed by varying the potential (pre-trained models or the reference \texttt{C-GAP-17}) used at different stages of the process: structure generation, and relaxation, application of strain, and evaluation of the elastic constants.

The excess energy of $a$-\ce{SiO2} is defined relative to the more stable $\alpha$-quartz crystalline phase (similar to previous analyses for $a$-Si \cite{roorda_structural_1991, deringer_realistic_2018}) and used here as an indicator of the quality of MLIP models, similar to Refs.~\citenum{erhard_modelling_2024} and \citenum{erhard_machine-learned_2022}. The excess energy was computed as an average over ten 510-atom configurations generated via the melt--quench procedure described above. The structures of $\alpha$-quartz and the amorphous samples were relaxed using the BFGS algorithm, allowing both cell vectors and atomic positions to vary, with a force convergence criterion of 0.01~eV/\AA{}. As in the Young’s modulus predictions, this protocol was systematically decomposed by varying the potential (pre-trained models or the reference \texttt{SiO$_{x}$-ACE-24}) used at different stages of the process: structure generation, and relaxation and energy evaluation.

\subsection*{Reference potentials}
For structure generation and application tests, model performance was evaluated against established specialist MLIPs: \texttt{C-GAP-17} for amorphous carbon \cite{deringer_machine_2017}, \texttt{SiO$_x$-ACE-24} for amorphous silica \cite{erhard_modelling_2024}, and \texttt{GST-ACE-24} for amorphous Ge--Sb--Te \cite{zhou_full-cycle_2025}. These potentials provide reference points for the corresponding amorphous phases and have been validated in previous studies against structural and thermodynamically relevant properties \cite{deringer_machine_2017, erhard_modelling_2024, zhou_full-cycle_2025}. Both GAP \cite{erhard_machine-learned_2022, zhou_device-scale_2023} and ACE \cite{erhard_modelling_2024, zhou_full-cycle_2025} specialist potentials were available for silica and GST systems; in this work, ACE-based models were selected as the reference model due to their computational efficiency and the use of augmented training datasets with broader configurational coverage compared to their GAP alternatives.

\subsection*{Pre-trained MLIPs}
For the numerical benchmark (Fig.~2), all pre-trained MLIPs were evaluated in a zero-shot setting using either the inference interfaces provided by their respective authors or, where available, the corresponding implementations in the \texttt{graph-pes} package \cite{gardner_graph-pes_2024}.

For the structure generation and application tests, we focus on a series of MLIP models from the MACE family \cite{batatia_foundation_2025} (available at \url{https://github.com/ACEsuit/mace}). Specifically, the models were: \texttt{MACE-MP-0b3}, \texttt{MACE-MPA-0}, \texttt{MACE-OMAT-0}, \texttt{MACE-MATPES-PBE-0}, and \texttt{MACE-MH-1}. The results in Figures 3--5 are based on zero-shot evaluations of these models.

\subsection*{Fine-tuning}
\subsubsection*{Fine-tuning datasets}
Independent fine-tuning datasets were generated for amorphous carbon, silica, and Ge--Sb--Te using the same melt--quench protocols and reference models employed to construct the corresponding AM26 subsets (\texttt{C-GAP-17} \cite{deringer_machine_2017}, \texttt{SiO$_x$-ACE-24} \cite{erhard_modelling_2024}, and \texttt{GST-ACE-24} \cite{zhou_full-cycle_2025}, respectively). These structures were generated independently of AM26 to ensure complete separation between the fine-tuning and benchmarking datasets.

For $a$-C, 100 structures were generated at each density between 1.5--3.5~g/cm$^{-3}$. For $a$-\ce{SiO2}, 125 structures were generated at each melt--quench rate (10$^{11}$, 10$^{12}$, 10$^{13}$, and 10$^{14}$~K/s). For $a$-GST, 75 structures were generated at each of seven compositions spanning the GeTe--\ce{Sb2Te3} tie-line.

Each dataset was subsequently divided into training and validation sets while preserving the corresponding density, quench-rate, or compositional distributions. All structures were labelled with DFT energies and forces using the same workflow as employed for AM26.

\subsubsection*{Training protocol}
Fine-tuning of the \texttt{MACE-MP-0b3} model (Fig.~6) was carried out using the \texttt{graph-pes} package \cite{gardner_graph-pes_2024} following a ``naive'' protocol, in which pre-trained weights are updated directly using the fine-tuning dataset. All models were trained on an NVIDIA RTX A6000 GPU in \texttt{float32} precision. A 6~\AA{} cut-off with a learnable offset was used. Training employed a learning rate of 0.0001, AdamW optimizer \cite{loshchilov_decoupled_2019}, and a batch size of 5. The loss function combined energy and force terms in a 1:10 ratio, mimicking the original training procedure in Ref.~\citenum{batatia_foundation_2025}. Training was terminated when the validation loss did not decrease for 50 consecutive epochs. The resulting fine-tuned model was evaluated using the \texttt{graph-pes} interface with ASE \cite{hjorth_larsen_atomic_2017} for evaluation and with LAMMPS \cite{thompson_lammps_2022} to run MD simulations. 

Learning curves were constructed using nested subsampling of the available training data (Fig.~6a). The fine-tuning dataset comprised 400 structures, with 100 structures from each quench rate (10$^{11}$, 10$^{12}$, 10$^{13}$, and 10$^{14}$~K/s). For each repeat, the dataset was randomly shuffled and training subsets of increasing size ($n = 4$, 8, 24, 40, 80, 100, 200, 300, and 400) were drawn without replacement, while maintaining equal representation from each quench rate. A model was fine-tuned for each subset size using a common independent validation set.

This procedure was repeated three times using different random seeds for the initial shuffle, and the reported values correspond to the mean and standard deviation across repeats. Owing to the nested-sampling strategy, overlap between training subsets increases with subset size. Consequently, the reported standard deviations at smaller $n$ reflect both variation in the selected training structures and training stochasticity, whereas at larger $n$ they increasingly reflect training stochasticity alone.

Additional analyses were performed to assess the influence of fine-tuning dataset composition (Fig.~6b). For each individual quench rate (10$^{11}$, 10$^{12}$, 10$^{13}$, and 10$^{14}$~K/s), training subsets of increasing size ($n = 1$, 2, 5, 10, 25, 50, 75, and 100) were constructed using nested subsampling without replacement from the corresponding pool of 100 structures. The same random shuffles used for the learning curve analysis were employed, and results were averaged over three independent repeats. 

\clearpage
\section*{Supplementary Figures}
\begin{figure*}[htbp]
    \centering
    \includegraphics[width=0.96\linewidth]{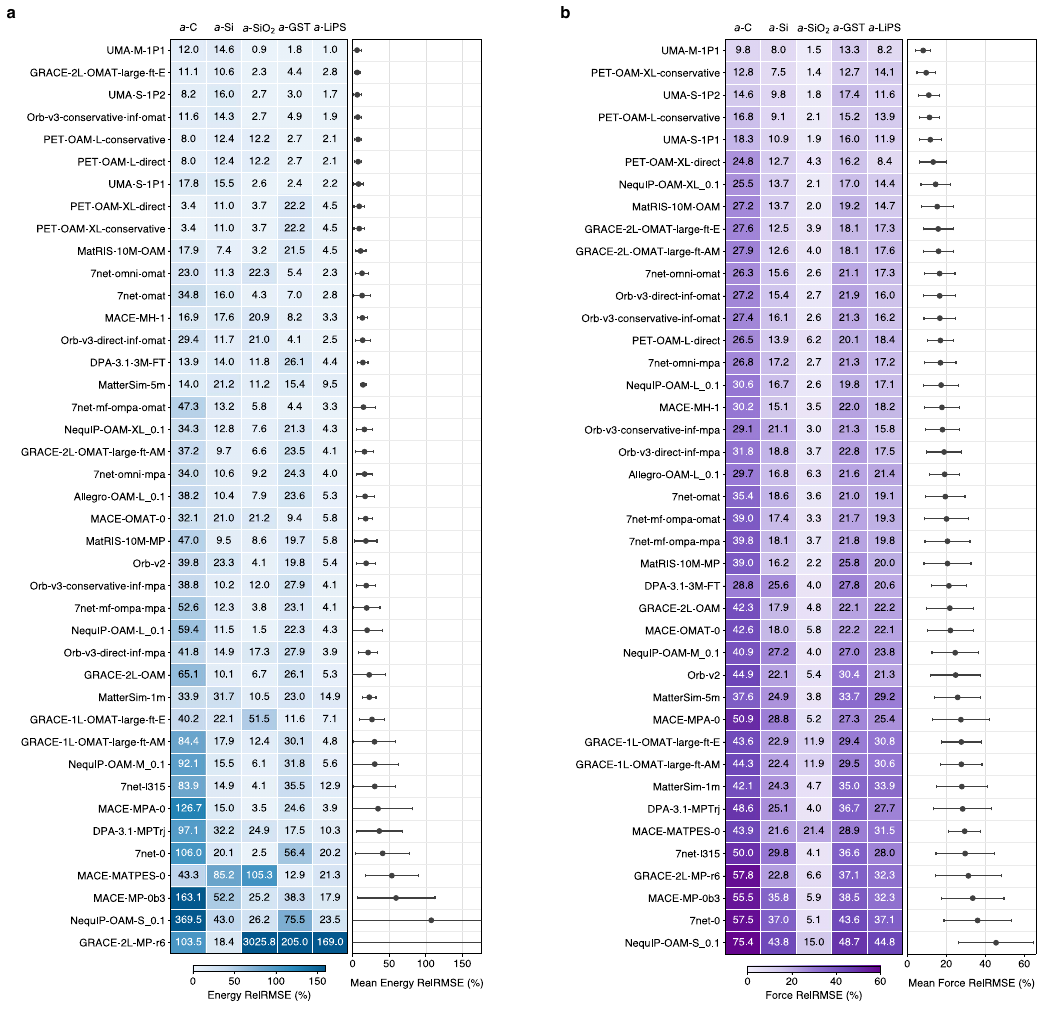}
    \caption{\textbf{Extended numerical results for evaluations on AM26} {\em (as a supplement to Fig.~2 in the main text)}.
    Energy performance is shown in panel (\textbf{a}), and force performance in panel (\textbf{b}). Heatmaps (\textit{left}) show per-system relative root-mean-square errors (RelRMSEs). Dot plots (\textit{right}) report the mean RelRMSE across the five AM26 subsystems, with error bars indicating the standard deviation across systems. Models are ranked by mean RelRMSE for the respective target.} 
    \label{fig:figure_s1}
\end{figure*}

{\em Note 1:} We note that MatterSim models are referred to with an abbreviated name; for example, \texttt{MatterSim-1M} refers to the \texttt{MatterSim-v1.0.0-1M} model.

{\em Note 2:} Figure~\ref{fig:figure_s1} extends the numerical benchmark in Fig.~2 to all 41 pre-trained models evaluated in this work. Several additional trends become apparent from this broader comparison. Performance on AM26 generally improves with increasing model scale across the UMA, MatterSim, and NequIP families (Fig.~S1) \cite{wood_uma_2026, yang_mattersim_2024, tan_high-performance_2025}, suggesting that increased representational capacity helps generalisation to the amorphous state. PET-OAM \cite{mazitov_pet-mad_2025} models form a notable exception: while the \texttt{-XL} variants outperform the corresponding \texttt{-L} models in force prediction, the opposite trend is observed for energies. For architectures where both direct and conservative formulations exist (PET-OAM and Orb \cite{mazitov_pet-mad_2025, neumann_orb_2024, rhodes_orb-v3_2025}), no consistent advantage is observed: conservative PET-OAM models outperform their direct counterparts in force prediction, whereas Orb-conservative models achieve lower energy errors; otherwise, performance differences between formulations are small. A clearer trend emerges with respect to pre-training data composition, where models trained on more diverse datasets such as OMat24 and MPA outperform those trained solely on MPTrj \cite{barros-luque_open_2026, schmidt_improving_2024, deng_chgnet_2023}; this trend is particularly apparent within the MACE family \cite{batatia_foundation_2025, batatia_cross_2025}.

\clearpage
\begin{figure*}[htbp]
    \centering
    \includegraphics[width=\linewidth]{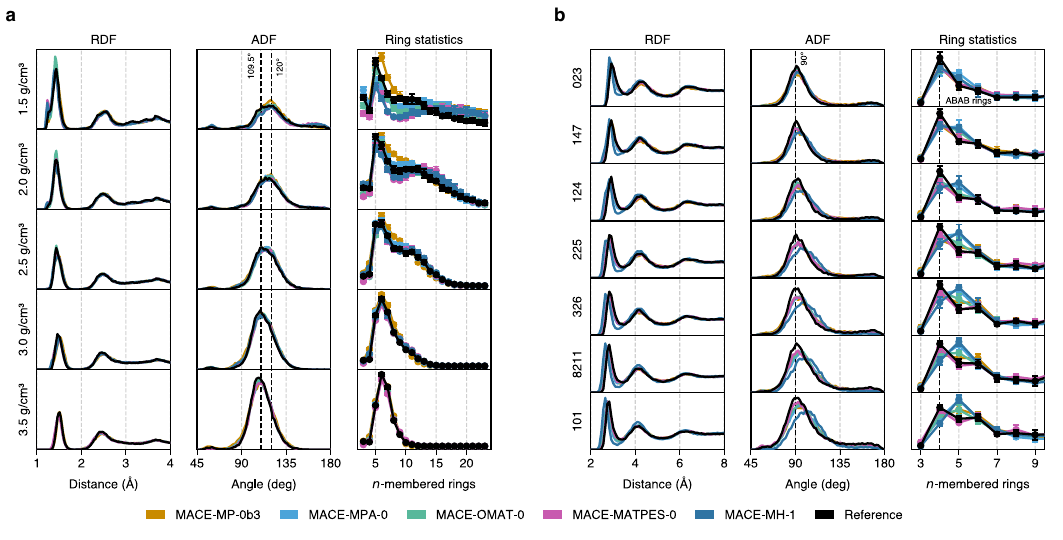}
    \caption{\textbf{Extended structural analysis for amorphous carbon and GST.} {\em (as a supplement to Fig.~3 in the main text)}.
    Extended results corresponding to the representative cases shown in the main text.
    (\textbf{a}) Amorphous carbon structures across the density range of 1.5--3.5 g/cm$^{3}$, generated by melt--quench simulations driven by pre-trained MLIPs, and compared against structures obtained using the \texttt{C-GAP-17} reference potential \cite{deringer_machine_2017}. 
    (\textbf{b}) Amorphous GST compositions along the \ce{GeTe}--\ce{Sb2Te3} tie-line generated by melt--quench simulations driven by pre-trained MLIPs, and compared against structures obtained using the \texttt{GST-ACE-24} reference potential \cite{zhou_full-cycle_2025}. In both (\textbf{a}) and (\textbf{b}), results shown are an average over ten amorphous structures generated at each density or composition.}
    \label{fig:figure_s2}
\end{figure*}

\clearpage

\begin{figure*}[htbp]
    \centering
    \includegraphics[width=\linewidth]{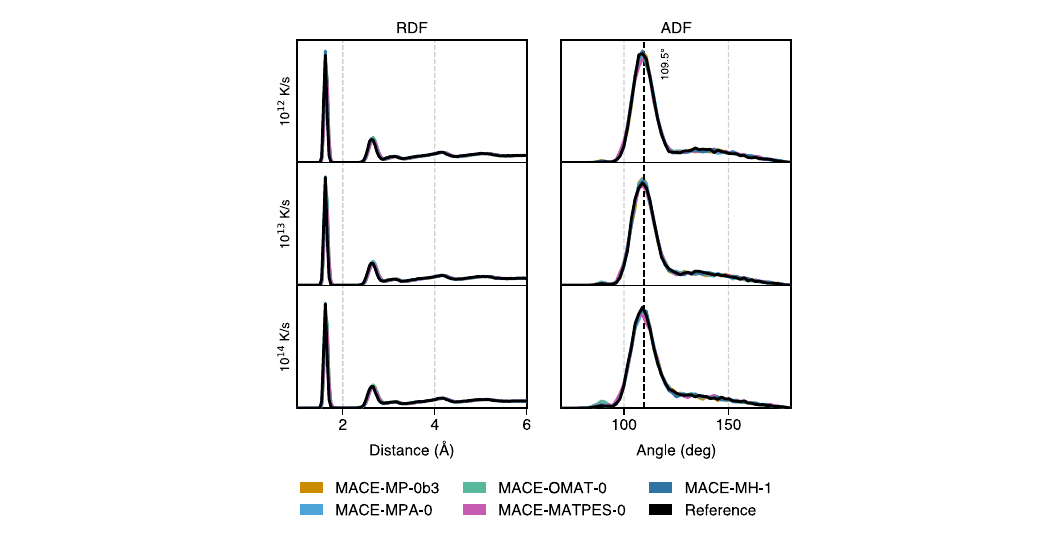}
    \caption{\textbf{RDFs and ADFs for amorphous silica.}
    Amorphous silica structures generated by melt--quench simulations with varying quench rates (10$^{12}$, 10$^{13}$, and 10$^{14}$ K/s)  driven by pre-trained MLIPs, and compared to structures obtained using the \texttt{SiO$_x$-ACE-24} reference potential \cite{erhard_modelling_2024}. Results shown are an average over ten amorphous structures generated with each quench rate.}
    \label{fig:figure_s2}
\end{figure*}

\clearpage
\begin{figure*}[htbp]
    \centering
    \includegraphics[width=\linewidth]{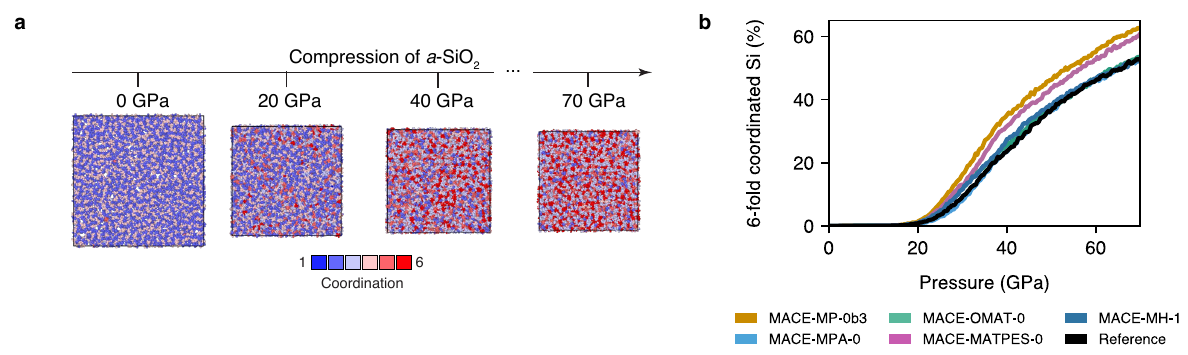}
    \caption{\textbf{Compression behaviour of amorphous silica.}
    (\textbf{a}) Compression of amorphous silica between 0--70~GPa simulated using \texttt{MACE-MP-0b3}. Representative snapshots are coloured according to the coordination number of Si atoms. 
    (\textbf{b}) Fraction of sixfold-coordinated Si atoms as a function of pressure for the reference \texttt{SiO$_x$-ACE-24} moel and pre-trained models. Results are averaged over three independent compression trajectories.}
    \label{fig:figure_s4}
\end{figure*}

{\em Note:} Figure \ref{fig:figure_s4} compares the pressure-induced coordination change predicted by pre-trained models with the reference \texttt{SiO$_x$-ACE-24} model \cite{erhard_modelling_2024}. All models reproduce the onset of pressure-induced densification and the overall increase in the fraction of sixfold-coordinated Si atoms with pressure. Although some pre-trained models overestimate the final high-pressure coordination fraction by up to $\sim$10\%, the overall compression behaviour remains well captured.

\clearpage
\begin{figure*}[htbp]
    \centering
    \includegraphics[width=\linewidth]{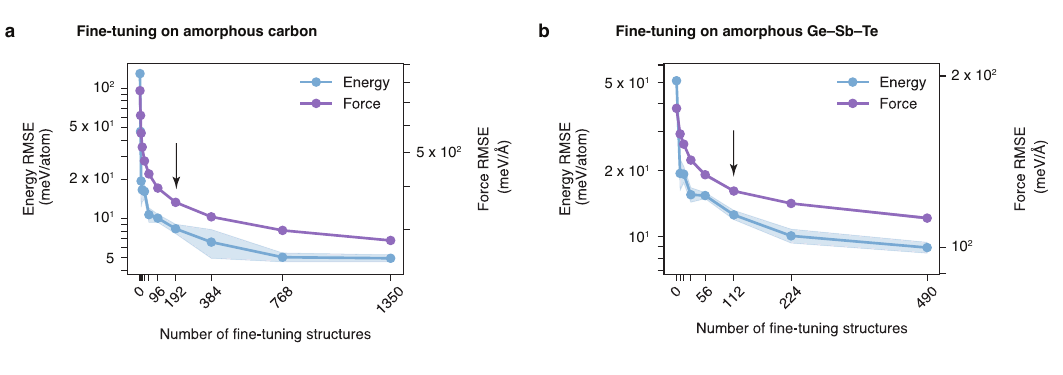}
    \caption{\textbf{Fine-tuning on amorphous carbon and Ge--Sb--Te.} Learning curves showing the performance of \texttt{MACE-MP-0b3} as a function of fine-tuning dataset size for (\textbf{a}) amorphous carbon and (\textbf{b}) amorphous GST. Points represent the mean over three independent repeats, with shaded regions indicating a standard deviation. Arrows denote the fine-tuned models selected for subsequent downstream simulations, corresponding to the onset of diminishing returns in energy and force prediction errors.
    }
    \label{fig:figure_s5}
\end{figure*}

\clearpage
\begin{figure*}[htbp]
    \centering
    \includegraphics[width=\linewidth]{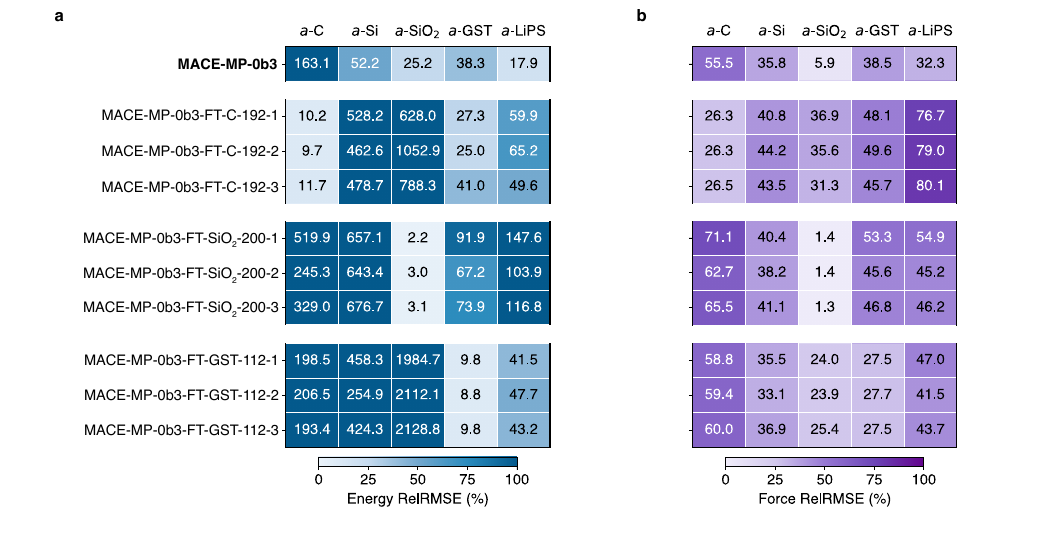}
    \caption{\textbf{Cross-domain performance of fine-tuned models across the AM26 benchmark.}
    Heatmaps show (\textbf{a}) energy and (\textbf{b}) force RelRMSE values across the five AM26 subsystems for the zero-shot \texttt{MACE-MP-0b3} model and fine-tuned variants. The zero-shot model is shown at the top, followed by groups of models fine-tuned on amorphous carbon, silica, and GST datasets. Within each domain, three independently fine-tuned models are shown, each trained on the same number of amorphous structures but using different randomly sampled fine-tuning datasets. Fine-tuned models are labelled according to the convention \texttt{MACE-MP-0b3-FT-X-N-R}, where \texttt{X} denotes the target system (\texttt{C}, \texttt{\ce{SiO2}}, or \texttt{GST}), \texttt{N} is the number of fine-tuning structures (see Fig.~S5), and \texttt{R} identifies the independent fine-tuning repeat.}
    \label{fig:figure_s6}
\end{figure*}

{\em Note:} Fine-tuning generally improves accuracy within the target domain, but often leads to substantial degradation in performance across non-target systems, consistent with catastrophic forgetting.

\clearpage
\begin{figure*}[htbp]
    \centering
    \includegraphics[width=\linewidth]{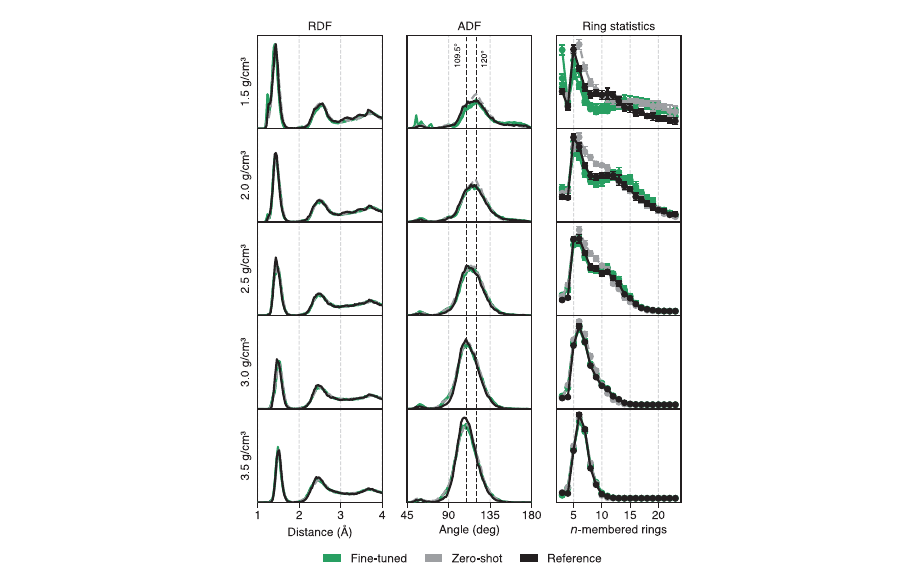}
    \caption{\textbf{Structural analysis of amorphous carbon generated using fine-tuned MACE-MP-0b3 models.}  {\em (as a supplement to Fig.~6 in the main text)}. 
    Radial distribution functions (RDFs), angular distribution functions (ADFs), and ring statistics for amorphous carbon structures across the density range 1.50--3.50 g/cm$^3$. Results from the three independently fine-tuned MACE-MP-0b3 models are compared with the corresponding zero-shot model and the reference C-GAP-17 model \cite{deringer_machine_2017}.}
    \label{fig:figure_s7}
\end{figure*}

\clearpage
\begin{figure*}[htbp]
    \centering
    \includegraphics[width=\linewidth]{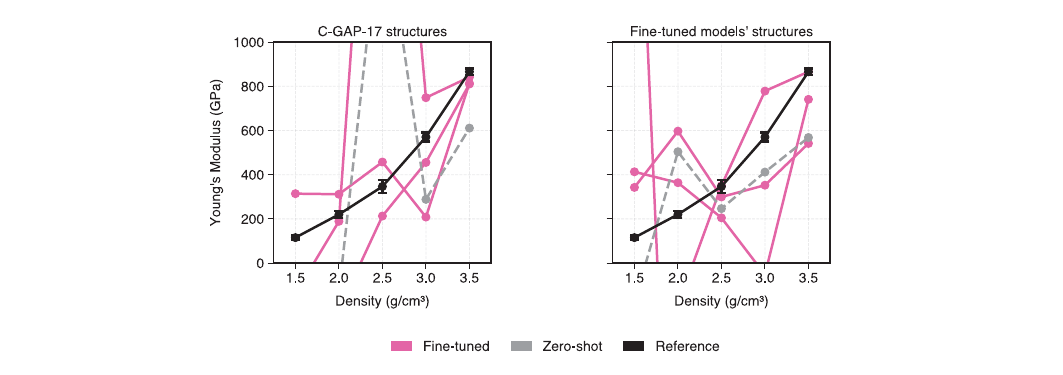}
    \caption{\textbf{Young's modulus of amorphous carbon as a function of density, comparing zero-shot and fine-tuned MACE models.}
    In the left column, Young's moduli are evaluated on reference \texttt{C-GAP-17} structures \cite{deringer_machine_2017}; in the right column, they are evaluated on structures generated by the corresponding zero-shot or fine-tuned model. The black reference curve, corresponding to \texttt{C-GAP-17} evaluated on its own structures, is shown in both columns for comparison. Mean values over 10 independent $a$-C structures are displayed, and error bars indicate one standard deviation. Standard deviations are omitted where their magnitude would obscure the underlying trends, and lines are included as guides to the eye to facilitate comparison with the reference.}
    \label{fig:figure_s8}
\end{figure*}

{\em Note:} Figure~\ref{fig:figure_s8} compares the Young's modulus predicted by the zero-shot and fine-tuned \texttt{MACE-MP-0b3} models. Fine-tuning yields only modest improvements: unphysical negative Young's moduli become less frequent, but the expected density dependence remains poorly reproduced. Similar behaviour is observed regardless of whether the elastic properties are evaluated on reference \texttt{C-GAP-17} structures or structures generated by the corresponding zero-shot or fine-tuned model, indicating that the dominant limitation remains the learned potential-energy surface rather than the generated amorphous configurations.

This limited improvement is perhaps unsurprising given that the fine-tuning datasets used in the present work comprise only equilibrium amorphous configurations and contain no explicit information about the response of the potential-energy surface to mechanical deformation. In contrast, Ref.~\citenum{gao_benchmarking_2026} demonstrated that fine-tuning on datasets containing strained crystalline configurations substantially improves the prediction of crystalline elastic properties. Extending fine-tuning datasets to include strained amorphous configurations may therefore provide a more effective route to improving elastic-property predictions, although such approaches lie beyond the scope of the present work.

\clearpage
\begin{figure*}[htbp]
    \centering
    \includegraphics[width=\linewidth]{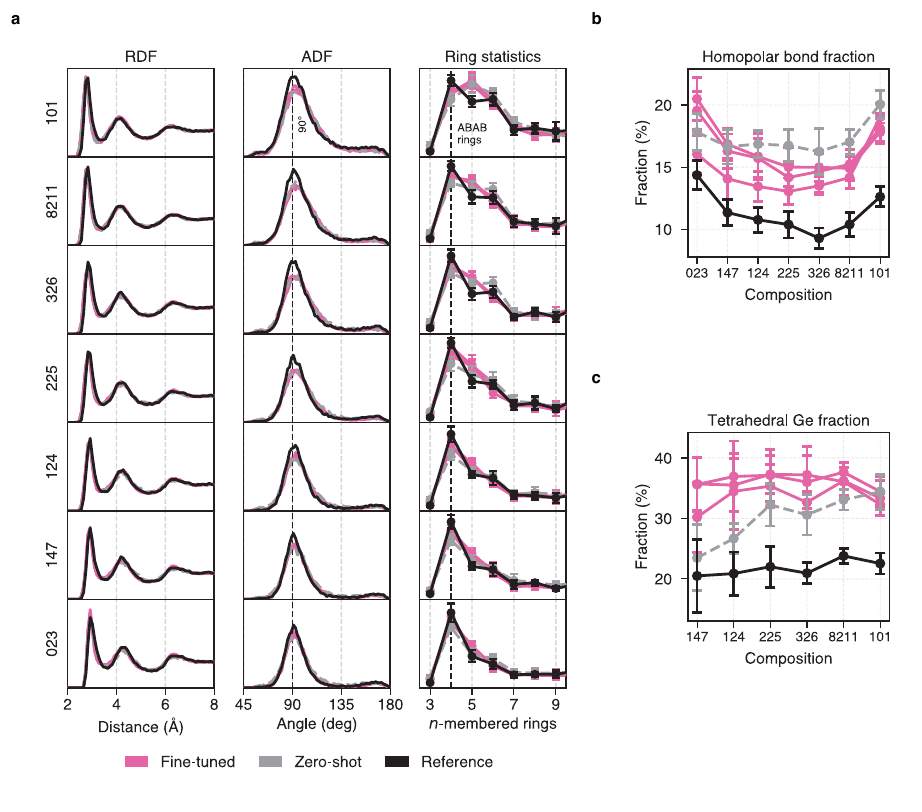}
    \caption{\textbf{Structural analysis of amorphous GST generated using fine-tuned MACE-MP-0b3 models.}  {\em (as a supplement to Fig.~6 in the main text)}. 
    (\textbf{a}) Radial distribution functions (RDFs), angular distribution functions (ADFs), and ring statistics for amorphous GST structures across the GeTe--\ce{Sb2Te3} tie-line. (\textbf{b}) Homopolar bond counts across the tie-line. (\textbf{c}) Tetrahedral Ge populations across the tie-line. Results from the three independently fine-tuned MACE-MP-0b3 models are compared with the corresponding zero-shot model and the reference GST-ACE-24 model \cite{zhou_full-cycle_2025}.}
    \label{fig:figure_s9}
\end{figure*}

\clearpage
\section*{Supplementary Tables}

\begin{table}[htbp]
    \centering
    \caption{Fraction of structures of each MACE pre-trained model lying within the 95\% confidence ellipses of the \texttt{C-GAP-17} reference manifold of each density (1.5--3.5~g/cm$^3$). The last column shows the mean fraction of each model across all densities. The last row shows the mean fraction of each density across all models. Potentials are ordered from best to worst match to the reference.}
    \begin{threeparttable}
        \begin{tabular}{lccccc c}
            \toprule
            \textbf{Potential} & 
            \textbf{1.5} & 
            \textbf{2.0} & 
            \textbf{2.5} & 
            \textbf{3.0} & 
            \textbf{3.5} & 
            \textbf{Mean (\%)} \\
            & \textbf{g/cm$^{3}$} & \textbf{g/cm$^{3}$} & \textbf{g/cm$^{3}$} & \textbf{g/cm$^{3}$} & \textbf{g/cm$^{3}$} & \\
            \midrule
            \texttt{MACE-MPA-0}       & 3   & 86  & 95  & 87  & 17  & 57.6  \\ 
            \texttt{MACE-OMAT-0}      & 0   & 41  & 67  & 50  & 66  & 44.8  \\ 
            \texttt{MACE-MATPES-0}    & 0   & 6   & 81  & 72  & 56  & 43.0  \\ 
            \texttt{MACE-MH-1}        & 1   & 50  & 40  & 30  & 75  & 39.2  \\ 
            \texttt{MACE-MP-0b3}      & 0   & 0   & 5   & 0   & 0   & 1.0   \\ 
            \midrule
            \textbf{Mean} & 0.8 & 36.6 & 57.6 & 47.8 & 42.8 & 37.1 \\
            \bottomrule
        \end{tabular}
    \end{threeparttable}
\end{table}

\FloatBarrier
\section*{Supplementary References}